\documentclass[aps,prd,preprintnumbers,showpacs,superscriptaddress,nofootinbib,amsmath,amssymb,floats,floatfix,showkeys,notitlepage,longbibliography]{revtex4-1}
\usepackage{comment}
\usepackage{graphicx}
\usepackage{subfigure}
\usepackage{palatino}
\usepackage[commandnameprefix=always]{changes}
\usepackage{hyperref}
\hypersetup{colorlinks=true,linkcolor=red,urlcolor=red,citecolor=red}
\usepackage[toc,page]{appendix}
\usepackage[normalem]{ulem}

\usepackage{lipsum}
\usepackage{graphicx}
\usepackage{subfigure}
\usepackage{palatino}
\usepackage{float}
\usepackage{sans}
\usepackage{adjustbox}
\usepackage{latexsym}
\usepackage{amsmath}
\usepackage{amssymb}
\usepackage{amsfonts}
\usepackage{dcolumn}
\usepackage{bm}
\usepackage{tikz}
\usepackage{bigints}
\usepackage{array,tabularx,multirow,booktabs}
\usepackage[tracking=true]{microtype}
\SetTracking{}{500}
\SetTracking{encoding={*}, shape=sc}{40}
\allowdisplaybreaks
\usepackage{adjustbox}
\usepackage{latexsym}
\usepackage{amsmath}
\usepackage{amssymb}
\usepackage{amsfonts}
\usepackage{dcolumn}
\usepackage{bm}
\usepackage{tikz}
\usepackage{bigints}
\usepackage{array,tabularx,multirow,booktabs}
\usepackage[tracking=true]{microtype}
\usepackage{color}
\usepackage{hyperref}
\usepackage{tcolorbox}
\usepackage{changepage}
\usepackage{mdframed}
\usepackage{fancybox}
\usepackage{hyperref}
\usepackage{tcolorbox}
\usepackage{changepage}
\usepackage{mdframed}
\usepackage{tcolorbox}
\allowdisplaybreaks

\begin{document}

\title{Overcoming Barriers: Kramers' Escape Rate Analysis of Metastable Dynamics in First-Order Multi-Phase Transitions }

\author{Mohammad Ali S. Afshar}
\email{m.a.s.afshar@gmail.com}
\affiliation{Department of Physics, Faculty of Basic
Sciences, University of Mazandaran\\ P. O. Box 47416-95447, Babolsar, Iran}
\affiliation{School of Physics, Damghan University, P. O. Box 3671641167, Damghan, Iran}
\affiliation{Canadian Quantum Research Center, 204-3002 32 Ave Vernon, BC V1T 2L7, Canada}

\author{Saeed Noori Gashti}
\email{saeed.noorigashti@stu.umz.ac.ir; saeed.noorigashti70@gmail.com}
\affiliation{School of Physics, Damghan University, P. O. Box 3671641167, Damghan, Iran}

\author{Mohammad Reza Alipour}
\email{mr.alipour@stu.umz.ac.ir}
\affiliation{School of Physics, Damghan University, P. O. Box 3671641167, Damghan, Iran}
\affiliation{Department of Physics, Faculty of Basic
Sciences, University of Mazandaran\\ P. O. Box 47416-95447, Babolsar, Iran}

\author{Jafar Sadeghi}
\email{pouriya@ipm.ir}
\affiliation{Department of Physics, Faculty of Basic
Sciences, University of Mazandaran\\ P. O. Box 47416-95447, Babolsar, Iran}
\affiliation{School of Physics, Damghan University, P. O. Box 3671641167, Damghan, Iran}
\affiliation{Canadian Quantum Research Center, 204-3002 32 Ave Vernon, BC V1T 2L7, Canada}

\vspace{1.5cm}\begin{abstract}
The expanding application of classical thermodynamic methods to black hole physics has yielded significant advances in characterizing phase transition behavior. Among these approaches, thermodynamic analysis—particularly kinetic formulations like the Kramers escape rate—provides a robust framework for probing black hole phase transitions with minimal relativistic constraints. This study investigates the kinetics and dynamic evolution of first-order phase transitions in black holes exhibiting multiple critical points, employing a particle-based escape rate model. The distinct free energy landscapes inherent to multi-critical systems, which can simultaneously support multiple local minima under specific thermodynamic conditions (temperature and pressure) within a given reference frame, raise fundamental questions regarding transition pathways.  We rigorously assess whether the Kramers escape rate retains its predictive validity in these complex multi-minima systems, as established for conventional single-minimum configurations. Furthermore, we examine whether transitions proceed via a sequential, stepwise mechanism between adjacent minima, or if pathways exist that bypass intermediate states through direct descent to the global minimum.  Our analysis of black holes undergoing multiphase transitions reveals both parallels and significant deviations from single-transition models. Crucially, we demonstrate that the Kramers escape rate remains a quantitatively reliable indicator of first-order phase transitions in black holes, even within multi-critical frameworks. This approach offers deeper insights into the governing energetic landscapes and kinetic processes underlying these phenomena.
\end{abstract}

\date{\today}



\maketitle
\tableofcontents

\section{Introduction}
Although most black hole models developed to date remain primarily theoretical due to cosmic-scale constraints and current observational limitations, they cannot be dismissed outright as mere hypothetical constructs. Given the complex and often unknown field interactions that may arise under diverse astrophysical conditions, the possibility of their formation remains significant.
Furthermore, the fractal and repetitive nature observed in many structures of the universe suggests a scientific justification for extending familiar physical principles observed at smaller scales to larger or smaller dimensions, while carefully considering contextual constraints. Based on these premises, one of the most intriguing analogies developed over the past five decades is the gravitational behavior of black holes, which closely resembles that of a thermodynamic ensemble—an analogy that has led to the fascinating field of black hole thermodynamics.  
In classical thermodynamics, critical points play a central role in determining the behavior of a system. Similarly, early studies of black hole thermodynamics suggested that the  Hawking-Page phase transition \cite{1}, which represents an intrinsic dimensional transition within a black hole to preserve energy principles, closely resembles  van der Waals-type phase transitions. This analogy was further strengthened by the proposal to treat black hole mass as  enthalpy \cite{2}, particularly in Anti-de Sitter (AdS) spacetime, where the cosmological constant and its conjugate variable correspond to pressure and volume, respectively. Under this interpretation, phase transitions between large and small charged AdS black holes were predicted and extensively studied, revealing striking similarities to gas-liquid transitions in the van der Waals model \cite{3,4,5,6,7,8,8.1,8.2,8.3}.\\\\ 
In classical thermodynamics, various methods are employed to analyze phase behavior depending on system conditions. One of the most widely used approaches is examining critical points and analyzing temperature and free energy functions. Recognizable diagrammatic structures, such as the swallowtail diagram, provide reliable insight into system behavior.  
In black hole thermodynamics also, this method has traditionally been favored. However, as our understanding of black hole behavior has advanced, more refined approaches have emerged. One such method involves studying phase transitions using topological charge, which has shown considerable consistency with conventional approaches and has led to extensive investigations \cite{9,10,11,12,13,14,15,16,17,18,19,20,21,22,22a,22b,22c,22d,22e,22f,22g}. Nonetheless, a crucial missing aspect in many studies is the static nature of endpoint analyses, which assume fixed trajectories while neglecting time dynamics, kinetics, and the potential chaotic nature of transitions during evolution.  
Recently, this issue has garnered attention, and efforts have been made to classify phase transition behavior based on kinetic principles. In this approach, black hole phase transitions can be interpreted as resulting from thermal fluctuations.  
Recent studies indicate that such thermally driven phase transitions can be understood through solutions to the Smoluchowski equation \cite{23,24,25,26,27,28,29,30}, which is fundamentally a probabilistic Fokker-Planck equation describing the diffusion process of a system constrained by potential barriers. From the perspective of free energy—particularly Gibbs free energy—it has been observed that transitions between small and large black holes are influenced by temperature fluctuations and potential barrier heights within the free energy landscape. According to studies, Gibbs free energy acts as an effective potential, driving the black hole phase transition \cite{31}.  
With this foundation, researchers have recently focused on Gibbs energy landscapes and Fokker-Planck-derived equations to evaluate kinetic aspects of phase transitions, specifically by computing the mean first passage time through the effective potential for different black hole models \cite{32,33,34,35,36,37,38,39,40,41,42,43,44,45,46,47,48,49,50,51,52,53}.\\\\ In addition to utilizing "the mean first passage time" method, an alternative approach has been employed to investigate the dynamics of black hole phase transitions. While still rooted in Fokker-Planck probability equations and free energy—which serves as an effective potential governing phase transitions—this method offers a distinct perspective on first-order phase transitions(FPT).  
In this framework, an analogy is drawn between the transition from a small black hole to a large black hole under the influence of free energy (acting as an effective potential) and a set of particles within a potential well that, due to thermal fluctuations, may escape its boundary \cite{54,55,56,57,58}. 
For particles, escape rates are commonly analyzed using Kramers' escape rate equation, which we have adapted to describe first-order black hole phase transitions. By reformulating the escape rate equation in terms of the effective potential governing black hole transitions, we examine the dynamical evolution of these phase shifts.\\\\  
In previous studies, we dynamically tracked the frame-by-frame evolution of escape rates, demonstrating how their initiation, peak activity, and subsequent decline closely align with first-order phase transitions across various black hole models. These studies also showed that in a very small range at the end of the direct phase transition from a small to a large black hole, a reverse process of chance emerges in which the probability of this overcomes the direct transition. We stated there that, given the very small range of occurrence of the reverse transition, this transition can in a way act as a controller to prevent uncontrolled processes and an attempt to end the direct process \cite{59,60}.
This suggests that the analogy is scientifically robust and provides insights into the kinetics of black hole phase transitions. Additionally, independent studies have confirmed this correspondence for numerous black hole configurations, reinforcing the validity of this approach in studying phase transitions from both a kinetic and frame-by-frame perspective.  
But what distinguishes this work from prior research—and motivates the application of this method to first-order phase transitions again—is the encounter with "multi-critical black holes". 
Investigations have shown that under specific parametric conditions in various black hole models—including four-dimensional Einstein gravity coupled with nonlinear electrodynamics \cite{61}, multiply rotating Kerr-AdS black holes \cite{62}, and spherically symmetric Lovelock gravity black holes \cite{63}—certain black hole solutions exhibit multi-critical behavior.  
But why is studying these models intriguing and significant? \\ 
Due to the distinct free energy structure in multi-critical models, which can simultaneously contain multiple local minima depending on the frame and observational perspective, several fundamental questions arise 
\begin{center}
\shadowbox{\parbox{0.9\textwidth}{
1. Does Kramers' escape rate still maintain the same predictive power in models with multiple minima, as it does for single-minimum, single-transition systems?\\
2. In scenarios where multiple local minima exist at a specific temperature and pressure (within a specific frame), do transitions occur in a stepwise, ordered manner, or is there a possibility of bypassing intermediate minima for a faster descent toward the lowest energy state?}}
\end{center}
To better conceptualize this, consider a simple analogy:\\ Suppose marbles are released from the top of a staircase with similar but slightly varying initial energies. While some marbles descend step-by-step, others may bypass certain steps entirely, skipping intermediate transitions and reaching the lower levels more directly. \\ 
Similarly, during a black hole phase transition, energy distribution across the system may not be uniform at the onset. Thus, could black holes exhibit analogous behavior—transitioning towards the final minimum without necessarily passing through all intermediate local minima?\\
Accordingly, to find answers to such questions, we will consider a model of black holes with multi-critical behavior and analyze its first-order phase transition using Kramers' escape rate.  
\section{Preliminary Methodology}
In this section, we present a structured discussion on the essential logical concepts underpinning this study. To ensure clarity and coherence, the exposition is divided into two subsections, each focusing on a fundamental aspect of the theoretical framework. The first subsection examines Kramers’ escape rate, a pivotal concept in statistical physics that quantifies transition dynamics over potential barriers. We explore its mathematical formulation and its relevance to thermodynamic processes. The second subsection introduces the Free Energy Landscape, which provides a comprehensive visualization of stability and transition states within complex systems. Together, these concepts establish a robust foundation for understanding the interplay between stochastic dynamics and energetic constraints in physical systems.
\subsection{Kramers' escape rate}
Kramers’ escape rate describes the rate at which a trapped particle in a local minimum of a potential energy well escapes over the barrier due to thermal fluctuations, ultimately reaching a more stable state. \\The primary objective is to quantify the average number of successful escapes under the influence of thermal conditions and according to classical energy principles, this escape process is governed by the presence of either more stable local minimums or a global minimum.\\ However, when analyzing transition probabilities in the presence of such a potential barrier, two distinct scenarios must be carefully differentiated:\\
1. Initial thermal fluctuations leading to escape over the local maximum, corresponding to the escape rate.\\  
2. Particle behavior after crossing the maximum, associated with the diffusion rate.\\
The critical aspect of the escape rate is that it must depend solely on the damping coefficient (dissipation), noise intensity (temperature), and potential function shape (barrier height). Simply put, the escape rate is determined by barrier height, considering local minima and maxima and the conditions enabling fluctuations. Since escape is defined by the first passage across the barrier, potential features beyond the maximum should not influence its value.\\ 
\begin{figure}[H]
 \begin{center}
 \includegraphics[height=8.5cm,width=8.5cm]{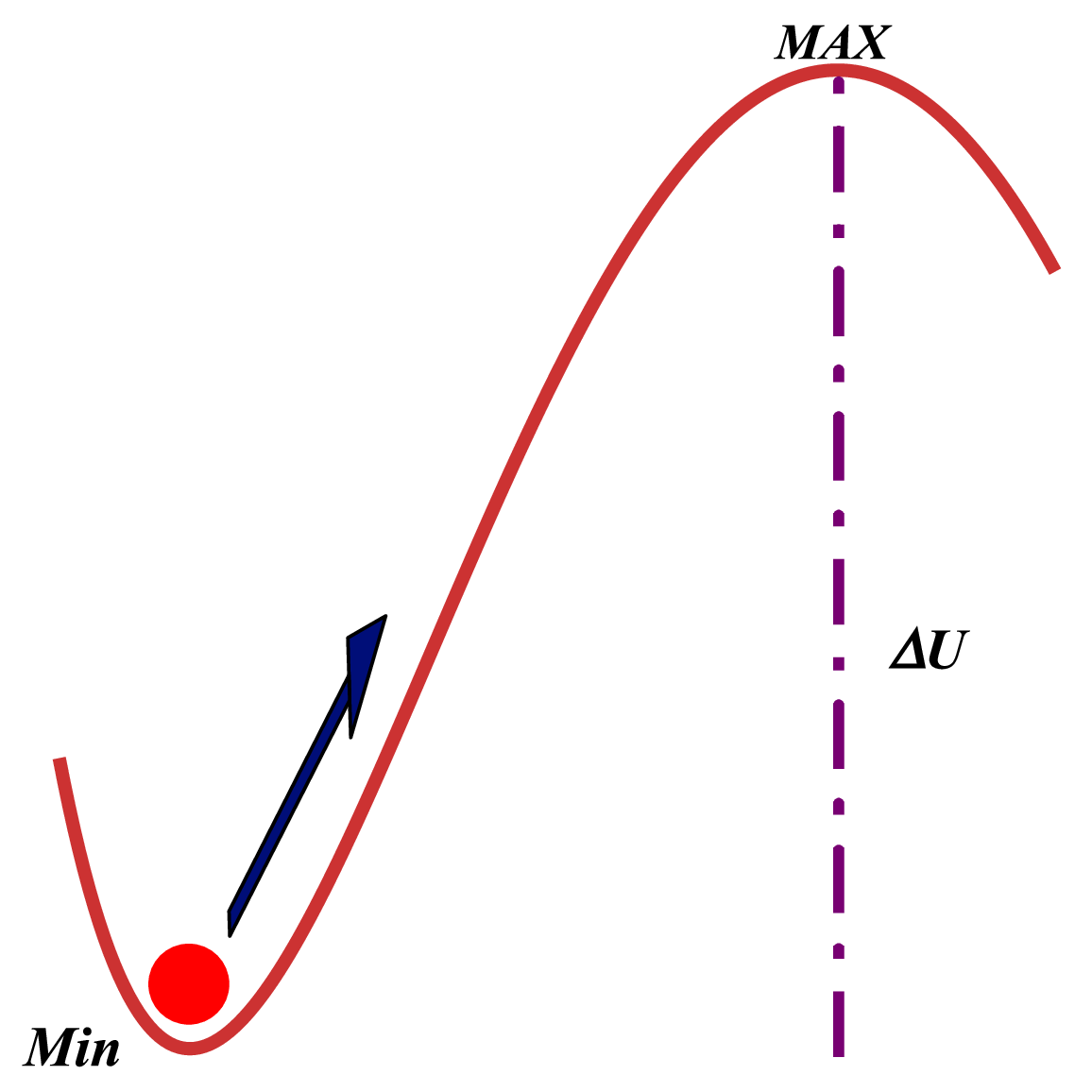}
 \caption{\small{The general form of potential U }}
 \label{m1}
\end{center}
\end{figure} 
To maintain generality while simplifying analysis, we adopt several assumptions.\\ As illustrated in Fig. (\ref{m1}), we assume that the system remains stable at the initial minimum (Min) and retains its equilibrium under small perturbations. To align with future studies, we consider only one escape direction, likely occurring via the nearest local maximum (Max).\\ 
Beyond this Max, as previously noted, additional local or global minima may exist—or the potential may be unbounded below—which affects diffusion speed but remains irrelevant to the escape rate. \\
When thermal baths are introduced into the system, random forces generated by thermal fluctuations displace the particle from its original position, eventually driving it over the potential barrier.\\ However, at the same time, damping forces act to slow the particle down, making a return to equilibrium increasingly difficult.\\\\ Based on this, three distinct damping regimes can be analyzed:\\
1. Low Damping: 
\begin{quote}
\setlength{\leftskip}{1mm}
a- The particle experiences minimal frictional forces.\\  
b- The escape rate is governed by energy diffusion, where particle inertia plays a dominant role.\\\  
c- Before escaping, the particle oscillates multiple times near the boundary of the potential well. \\ 
d- Due to weak dissipation, the escape rate remains relatively low.\\  
\end{quote} 
2. Intermediate Damping:
\begin{quote}
\setlength{\leftskip}{1mm} 
a- A more complex regime, balancing inertial effects with damping forces.\\  
b- Oscillatory behavior is reduced compared to the low-damping case, but is still present.\\ 
c- The escape rate remains moderate, but is not significantly increased.\\
\end{quote}   
3. Overdamped Regime:
\begin{quote}
\setlength{\leftskip}{1mm}  
a- Strong damping forces override inertial effects.\\  
b- The particle slowly drifts away from the potential well. \\ 
c- Here, thermal fluctuations primarily dictate escape, allowing the system to overcome the barrier.\\  
d- The escape process resembles slow diffusion over the barrier, leading to a significantly higher escape rate compared to previous regimes.\\
\end{quote} 
Since we aim to simulate black hole phase transition behavior, the overdamped case appears to be the most suitable. This is because:
\begin{quote}
\setlength{\leftskip}{1mm}
1. Fluctuating forces dominate over inertia, mirroring the influence of thermal variations in phase transitions.\\  
2. Due to enhanced combined effects, the system loses stability, increasing the likelihood of the particle escaping from the potential well.\\
\end{quote} 
The probability distribution \( \mathbf{P}(x, t) \) is governed by the Fokker-Planck equation \cite{59,60,64,65}:
\begin{equation}\label{k1}
\frac{\partial \mathbf{P}(x, t)}{\partial t} = - \frac{\partial \mathbf{J}(x, t)}{\partial x},
\end{equation}
where the probability current \( \mathbf{J}(x, t) \) is defined as:
\begin{equation}\label{k2}
\mathbf{J} = - \left( \frac{\partial U}{\partial x} \right) \frac{\mathbf{P}(x, t)}{\mathbf{\gamma}} - \mathbf{D} \frac{\partial \mathbf{P}(x, t)}{\partial x}.
\end{equation}
Here, \( U(x) \) represents the potential field, \(\gamma\) is the friction coefficient, and \( \mathbf{D} \) is the diffusion coefficient. The current can be rewritten as:
\begin{equation}\label{k3}
\mathbf{J} = - \mathbf{D} e^{-U / k_B T} \frac{\partial}{\partial x} \left( e^{-U / k_B T} \mathbf{P} \right),
\end{equation}
where \( k_B \) is the Boltzmann constant and \( T \) is the temperature. To integrate, we rearrange the equation:
\begin{equation}\label{k4}
\frac{\partial}{\partial x} \left( e^{-U / k_B T} \mathbf{P} \right) = - \frac{\mathbf{J} e^{U / k_B T}}{\mathbf{D}}.
\end{equation}
Upon integrating and considering the dominance of \( P(x_{\text{min}}) \) over \( x_{\text{max}} \), we obtain:
\begin{equation}\label{k5}
\mathbf{J} = \frac{\mathbf{D} e^{U(x_{\text{min}}) / k_B T} \mathbf{P}(x_{\text{min}})}{\int_{x_{\text{min}}}^{x_{\text{max}}} e^{-U(x) / k_B T} dx}.
\end{equation}
Applying an expansion around \( x_{\text{max}} \) and performing additional calculations yields:
\begin{equation}\label{k6}
\int_{x_{\text{min}}}^{x_{\text{max}}} e^{-U(x) / k_B T} dx \approx \sqrt{\frac{2\pi k_B T}{\left| \frac{d^2 U(x_{\text{max}})}{dx^2} \right|}} e^{U(x_{\text{max}}) / k_B T}.
\end{equation}
Defining \(\mathfrak{p}\) as the probability of a particle being inside the well, we derive:
\begin{equation}\label{k7}
\mathfrak{p} =\mathbf{P} \! \left(x_{\min}\right) \ \sqrt{\frac{2\pi  k_{B} T}{{| \frac{d^{2}}{d x^{2}}U( x_{min})| \! \left(x \right)|}}}.
\end{equation}
Using the definition of Kramers escape rate (\( K_{e-r} = \mathbf{J} / \mathbf{P} \)), we obtain:
\begin{equation}\label{k8}
 K_{e-r} = D \sqrt{\left| \frac{d^2 U(x_{\text{max}})}{dx^2} \frac{d^2 U(x_{\text{min}})}{dx^2}\right|} e^{-\frac{U(x_{\text{max}}) - U(x_{\text{min}})}{k_B T}} \frac{1}{2\pi k_B T}.
\end{equation}
Under thermal equilibrium conditions, \( D \) remains constant, allowing further simplifications:
\begin{equation}\label{k9}
 K_{e-r} = \sqrt{\left| \frac{d^2 U(x_{\text{max}})}{dx^2} \frac{d^2 U(x_{\text{min}})}{dx^2}\right|} e^{-\frac{U(x_{\text{max}}) - U(x_{\text{min}})}{D}} \frac{1}{2\pi}.
\end{equation}
This formulation applies when \( \Delta U \gg k_B T \), ensuring the escape process is thermally activated rather than spontaneous. 
\subsection{Free Eenergy Landscape}
The Gibbs free energy function plays a pivotal role in analyzing phase transitions within black hole thermodynamics, particularly under constant pressure conditions. Its dependence on enthalpy and temperature makes it an indispensable tool for exploring the stability and thermodynamic evolution of black hole configurations.
A notable graphical feature of Gibbs free energy is the swallowtail structure, which serves as an effective indicator of first-order phase transitions. Within the formalism of free energy analysis, Gibbs free energy is defined as:
\begin{equation}\label{k10}
G = H - T_{H} S,
\end{equation}
where \( H \) represents enthalpy, \( T_H \) is the Hawking temperature, and \( S \) denotes entropy. In the extended phase space of black hole thermodynamics, enthalpy is equated with mass (\( H \equiv M \))\cite{2}, allowing the expression to be rewritten accordingly.\\
One of the defining characteristics of a black hole is its temperature, particularly the Hawking temperature, which inherently allows for thermal fluctuations. During a phase transition, the system evolves from an initial locally stable state to a final globally stable configuration. In first-order phase transitions, black holes manifest in different size regimes—small, intermediate, and large—alongside transitional states.
A crucial point is that all final states must satisfy Einstein’s field equations. However, during the transition through these extreme configurations, the system may pass through unstable intermediate points, where strict adherence to Einstein’s equations is not necessarily required. Consequently, within a predefined energy range, black holes can follow multiple possible trajectories—corresponding to distinct “T” values that differ from the Hawking temperature ($T_{H}$)—between initial and final states, provided global stability is maintained \cite{31}.
Each of these spacetime configurations in the transition sequence is characterized by a distinct Gibbs free energy. This flexibility enables the incorporation of probabilistic and statistical methods, facilitating the development of a statistical framework for analyzing phase transitions. In such cases, black hole configurations can be effectively described using a generalized Gibbs free energy landscape\cite{31}:
\begin{equation}\label{k11}
G_L = M - T S,
\end{equation}
where \( G_L \) represents the generalized Gibbs free energy.
The temperature discrepancy between different configurations introduces the concepts of on-shell and off-shell thermodynamic states. An off-shell transition occurs at temperature \( T \), which does not necessarily correspond to a solution of Einstein’s equations. Only extremal points in the Gibbs free energy landscape indicate stable black hole phases that satisfy Einstein’s field equations.
\section{ The Model: 4D-Einstein-nonlinear electrodynamics Black hole}\label{Sec: The Model} 
In \cite{P1}, researchers introduce the first explicit examples of black hole multicritical points within the framework of non-linear electrodynamics in four-dimensional Einstein gravity. their findings indicate the existence of regions in the parameter space where multiple distinct black hole phases emerge under sufficiently large thermodynamic pressure.\\ Each transition between these phases is a first-order phase transition, with the associated coexistence lines terminating at unique second-order critical points.\\ As the pressure decreases to a particular threshold, all these distinct phases merge at a single multicritical point. Below this critical pressure, only two stable black hole phases remain, distinguished by their size and separated by a first-order phase transition.\\ To illustrate this phenomenon, consider the emergence of a quadruple point in black hole thermodynamics.\\\\ At high pressures, the system admits only a single black hole phase. As the pressure is gradually lowered:
\begin{quote}
\setlength{\leftskip}{1mm}
1. Two distinct black hole phases appear.\\
2. A third phase emerges, forming a triple point.\\
3. A fourth black hole phase arises, leading to a quadruple point.\\
4. At a specific tetra-critical pressure, all four phases merge into a single quadruple critical point.\\
\end{quote}
For pressures below this tetra-critical point, only two distinct phases persist: the largest and the smallest possible black holes, separated by a first-order phase transition. The findings suggest that black holes can exhibit multicomponent system behavior, analogous to condensed matter systems with multiple critical points \cite{P2}. Specifically: The formation of an n-tuple critical point in a charged black hole requires \( 2n - 1 \) conjugate pairs in Power-Maxwell theory. Gibbs Phase Rule dictates that there are n degrees of freedom associated with the multicritical point. The implications for the microstructure of black holes remain an open question, requiring further exploration.
Additionally, for certain thermodynamic parameters, researcher preserve up to 10 significant digits to ensure numerical precision, as even slight variations in these values can significantly alter phase behavior. Now, with respect to above concept, we want to study the keramers scape rate phase transion. so, in the extended thermodynamic framework of black hole chemistry, a fundamental relation governs the connection between the thermodynamic pressure \( P \) and the cosmological constant \( \Lambda \) as follows \cite{P1}:
\begin{equation}\label{M1}
P = -\frac{\Lambda}{8\pi G}, \quad \Lambda = -\frac{(D-1)(D-2)}{2l^2},
\end{equation}
where \( l \) represents the radius of the \( D \)-dimensional anti-de Sitter (AdS) space, and \( G \) is the Newton gravitational constant with dimensions. Here, we adopt the convention \( \hbar = c = 1 \). In this framework, the mass \( M \) of the black hole is interpreted as enthalpy rather than internal energy. The first law of black hole thermodynamics is expressed as:
\begin{equation}\label{M2}
\delta M = T \delta S + V \delta P + \phi \delta Q + \Omega \delta J,
\end{equation}
with the corresponding Smarr relation given by:
\begin{equation}\label{M3}
M = \frac{D-2}{D-3} (T S + \Omega J) + \phi Q - \frac{2}{D-3} P V.
\end{equation}
For a charged black hole in \( D \)-dimensional Einstein gravity, its entropy \( S \) and temperature \( T \) are related to its horizon area \( A \) and surface gravity \( \kappa \) as:
\begin{equation}\label{M4}
S = \frac{A}{4G}, \quad T = \frac{\kappa}{2\pi}.
\end{equation}
The thermodynamic volume \( V \), conjugate to the pressure \( P \), is defined by:
\begin{equation}\label{M5}
V = \left( \frac{\partial M}{\partial P} \right)_{S,Q,J}.
\end{equation}
Now, consider a general class of non-linear electrodynamics minimally coupled to four-dimensional Einstein gravity. The action governing this theory is formulated as \cite{P1, P3}:
\begin{equation}\label{M6}
S = \int d^4x \sqrt{-g} \left( R - 2\Lambda - \sum_{i=1}^{N} \alpha_i (F^2)^i \right).
\end{equation}
Here, the electromagnetic field strength tensor is defined as \( F_{\mu\nu} = \nabla_\mu A_\nu - \nabla_\nu A_\mu \), with its invariant \( F^2 = F_{\mu\nu} F^{\mu\nu} \). The coupling constants \( \alpha_i \) have dimensions \( [\alpha_i] = L^{2(i-1)} \), while \( A_\mu \) represents the gauge field corresponding to a \( U(1) \) symmetry. The standard Einstein-Maxwell theory is recovered by setting \( \alpha_1 = 1 \) and \( \alpha_i = 0 \) for \( i > 1 \). By defining the nonlinear Lagrangian density \( L_{EM} \) as:
\begin{equation}\label{M7}
L_{EM} = -\sum_{i=1}^{N} \alpha_i (F^2)^i,
\end{equation}
the equations governing the Einstein-Power-Maxwell system take the form:
\begin{equation}\label{M8}
G_{\mu\nu} = -2 \frac{d L_{EM}}{d F^2} F^\lambda_\mu F_{\nu\lambda} + \frac{1}{2} g_{\mu\nu} L_{EM},
\end{equation}
\begin{equation}\label{M9}
\nabla_\mu \left( \frac{d L_{EM}}{d F^2} F^{\mu\nu} \right) = 0.
\end{equation}
To explore black hole solutions, we adopt the ansatz:
\begin{equation}\label{M10}
ds^2 = -U(r) dt^2 + \frac{dr^2}{U(r)} + r^2 d\Omega^2_2, \quad A_\mu = [\Phi(r), 0, 0, 0].
\end{equation}
This yields the field equations:
\begin{equation}\label{M11}
(r(U(r) - 1))' + r^2 \Lambda - r^2 \sum_{n=1}^{N} \left( n - \frac{1}{2} \right) \alpha_n (-2 (\Phi')^2)^n = 0,
\end{equation}
\begin{equation}\label{M12}
\frac{1}{2} r^2 \sum_{n=1}^{N} n \alpha_n (-2 (\Phi')^2)^n - Q (\Phi') = 0.
\end{equation}
Here, \( Q \) is an integration constant representing the electric charge of the black hole, while primes denote derivatives with respect to \( r \). For \( \Lambda = 0 \), appropriate choices of \( \alpha_i \) allow for asymptotically flat black holes with multiple horizons. For asymptotically AdS black holes, where \( \Lambda = -3/l^2 \), the solutions generalize naturally by introducing \cite{P1}:
\begin{equation}\label{M13}
\Phi = \sum_{i=1}^{K} b_i r^{-i}, \quad U = 1 + \sum_{i=1}^{K} c_i r^{-i} + \frac{r^2}{l^2}.
\end{equation}
The field equations then yield the relations:
\begin{equation}\label{M14}
c_1 = -2M, \quad c_i = \frac{4Q}{i+2} b_{i-1}, \quad (i > 1).
\end{equation}
For \( \alpha_1 = 1 \), the coefficients take the specific form \cite{P1}:
\begin{equation}\label{M15}
b_1 = Q, \quad b_5 = \frac{4}{5} Q^3 \alpha_2, \quad b_9 = \frac{4}{3} Q^5 (4\alpha_2^2 - \alpha_3),
\end{equation}
\begin{equation}\label{M16}
b_{13} = \frac{32}{13} Q^7 (24\alpha_2^3 - 12\alpha_3 \alpha_2 + \alpha_4),
\end{equation}
\begin{equation}\label{M17}
b_{17} = \frac{80}{17} Q^9 (176\alpha_2^4 - 132\alpha_2^2 \alpha_3 + 16\alpha_4 \alpha_2 + 9\alpha_3^2 - \alpha_5),
\end{equation}
\begin{equation}\label{M18}
b_{21} = \frac{64}{7} Q^{11} (1456\alpha_2^5 + 234\alpha_3^2 \alpha_2+ 208\alpha_4 \alpha_2^2 - 24\alpha_4 \alpha_3 - 1456\alpha_2^3 \alpha_3 - 20\alpha_5 \alpha_2 + \alpha_6).
\end{equation}
\begin{equation*}\label{(0)}
\beta =13056 \alpha_{2}^{6}-16320 \alpha_{2}^{4} \alpha_{3}+2560 \alpha_{2}^{3} \alpha_{4}+4320 \alpha_{2}^{2} \alpha_{3}^{2}-300 \alpha_{5} \alpha_{2}^{2}-720 \alpha_{2} \alpha_{3} \alpha_{4}-135 \alpha_{3}^{3}
\end{equation*}
\begin{equation}\label{M19}
b_{25}=\frac{448 \left(24 \alpha_{6} \alpha_{2}+30 \alpha_{3} \alpha_{5}+16 \alpha_{4}^{2}+\beta -\alpha_{7}\right) q^{13}}{25}
\end{equation}
Setting all coupling coefficients \( \alpha_i \) to zero for \( i > 1 \) simplifies the field equations, leading to the familiar Reissner-Nordström AdS black hole solution \cite{P1}:
\begin{equation}\label{M20}
\Phi = \frac{Q}{r}, \quad U(r) = \frac{r^2}{l^2} + 1 - \frac{2M}{r} + \frac{Q^2}{r^2}.
\end{equation}
This black hole possesses two distinct horizons, determined by appropriate choices of mass \( M \) and charge \( Q \). If instead certain nonzero coefficients \( \alpha_i \) are retained for \( i \leq n - 1 \), the number of distinct horizons can be extended to \( n \), reflecting the influence of nonlinear electrodynamic corrections. In this formulation, the parameter \( M \) represents the conserved charge associated with the timelike Killing vector \( \xi = \partial_t \) of the metric. To formalize this interpretation, we employ the Ashtekar-Das definition of conformal mass, a method designed to extract spacetime mass at the boundary through conformal regularization \cite{P4,P5}. A conformal transformation of the metric,
\begin{equation}\label{M21}
\bar{g}_{\mu\nu} = \bar{\Omega}^2 \, g_{\mu\nu},
\end{equation}
eliminates divergences in the asymptotic region (\( r \to \infty \)), providing a well-defined conserved charge. This charge is derived by integrating the conserved current:
\begin{equation}\label{M22}
Q(\xi) = \frac{\ell}{8\pi} \lim_{\Omega \to 0} \int\frac{\ell^2}{ \Omega} N^\alpha N^{\beta} \bar{C}^\nu_{\alpha\mu\beta} \xi^\nu d\bar{S}^\mu.
\end{equation}
Here, \( \bar{C}^\nu_{\alpha\mu\beta} \) is the Weyl tensor of the conformal metric. \( N_\mu = \partial_\mu \bar{\Omega} \) represents the normal to the boundary. \( d\bar{S}^\mu = \delta^\tau_\mu \ell^2 (d\cos\theta d\phi) \) is the spacelike surface element tangent to \( \bar{\Omega} = 0 \).
Despite the non-uniqueness of the conformal completion, the charge \( Q(\xi) \) remains independent of the chosen completion method. For convenience, we adopt \( \bar{\Omega} = \ell\Omega r^{-1} \), leading to the explicit charge expression \cite{P1}:
\begin{equation}\label{M23}
Q(\partial_\tau) = \lim_{r \to \infty} \left( M - \frac{Q^2}{r} - \frac{14b_5}{3r^5} - \frac{11b_9}{r^9} + \cdots + O \left( \frac{b_n}{r^n} \right) \right).
\end{equation}
Taking the asymptotic limit yields:
\begin{equation}\label{M24}
Q(\partial_\tau) = M.
\end{equation}
Thus, the conserved charge associated with the Killing vector \( \xi = \partial_t \) consistently recovers the black hole mass \( M \), affirming its thermodynamic interpretation. To illustrate the presence of a quadruple point, we analyze the conditions necessary for the emergence of multiple black hole horizons. Specifically, allowing nonzero values for the coupling coefficients \( \alpha_i \) up to \( i \leq 7 \) enables the formation of up to eight distinct horizons \cite{P3}. The scalar potential is given by \cite{P1}:
\begin{equation}\label{M25}
\Phi(r) = \frac{Q}{r} + \frac{b_5}{r^5} + \frac{b_9}{r^9} + \frac{b_{13}}{r^{13}} + \frac{b_{17}}{r^{17}} + \frac{b_{21}}{r^{21}} + \frac{b_{25}}{r^{25}},
\end{equation}
while the metric function takes the form:
\begin{equation}\label{M26}
U(r) = 1 - \frac{2M}{r} + \frac{Q^2}{r^2} + \frac{b_5 Q}{2r^6} + \frac{b_9 Q}{3r^{10}} + \frac{b_{13} Q}{4r^{14}} + \frac{b_{17} Q}{5r^{18}} + \frac{b_{21} Q}{6r^{22}} + \frac{b_{25} Q}{7r^{26}} + \frac{r^2}{l^2}.
\end{equation}
For simplicity, we set \( b_i = 0 \) for all \( i > 25 \). The Hawking temperature associated with the event horizon \( r_+ \) is given by \cite{P1}:
\begin{equation}\label{M27}
T = \frac{1}{4\pi r_+} \left( 1 + \frac{3r_+^2}{l^2} - \frac{Q^2}{r_+^2} - \frac{5b_5 Q}{2r_+^6} - \frac{3b_9 Q}{r_+^{10}} - \frac{13b_{13} Q}{4r_+^{14}} - \frac{17b_{17} Q}{5r_+^{18}} - \frac{7b_{21} Q}{2r_+^{22}} - \frac{25b_{25} Q}{7r_+^{26}} \right).
\end{equation}
The black hole entropy, thermodynamic volume, and pressure are given by:
\begin{equation}\label{M28}
S = \pi r_+^2, \quad V = \frac{4}{3} \pi r_+^3, \quad P = \frac{3}{8\pi l^2}.
\end{equation}
These quantities are expressed in Planckian units, where \cite{P6}:
\begin{equation}\label{M29}
l_P^2 = \frac{G \hbar}{c^3}.
\end{equation}
The equation of state governing the black hole thermodynamics is expressed as \cite{P1}:
\begin{equation}\label{M30}
P = \frac{T}{2r_+} - \frac{1}{8\pi r_+^2} + \frac{Q^2}{8\pi r_+^4} + \frac{5b_5 Q}{16\pi r_+^8} + \frac{3b_9 Q}{8\pi r_+^{12}} + \frac{13b_{13} Q}{32\pi r_+^{16}} + \frac{17b_{17} Q}{40\pi r_+^{20}} + \frac{7b_{21} Q}{16\pi r_+^{24}} + \frac{25b_{25} Q}{56\pi r_+^{28}}.
\end{equation}
The Gibbs free energy is given by:
\begin{equation}\label{M31}
G = M - T S.
\end{equation}
\section{Analyzing of the model's First-order phase transition }\label{Sec: The Model}
\begin{figure}[H]
 \begin{center}
 \subfigure[]{
 \includegraphics[height=6cm,width=5.6cm]{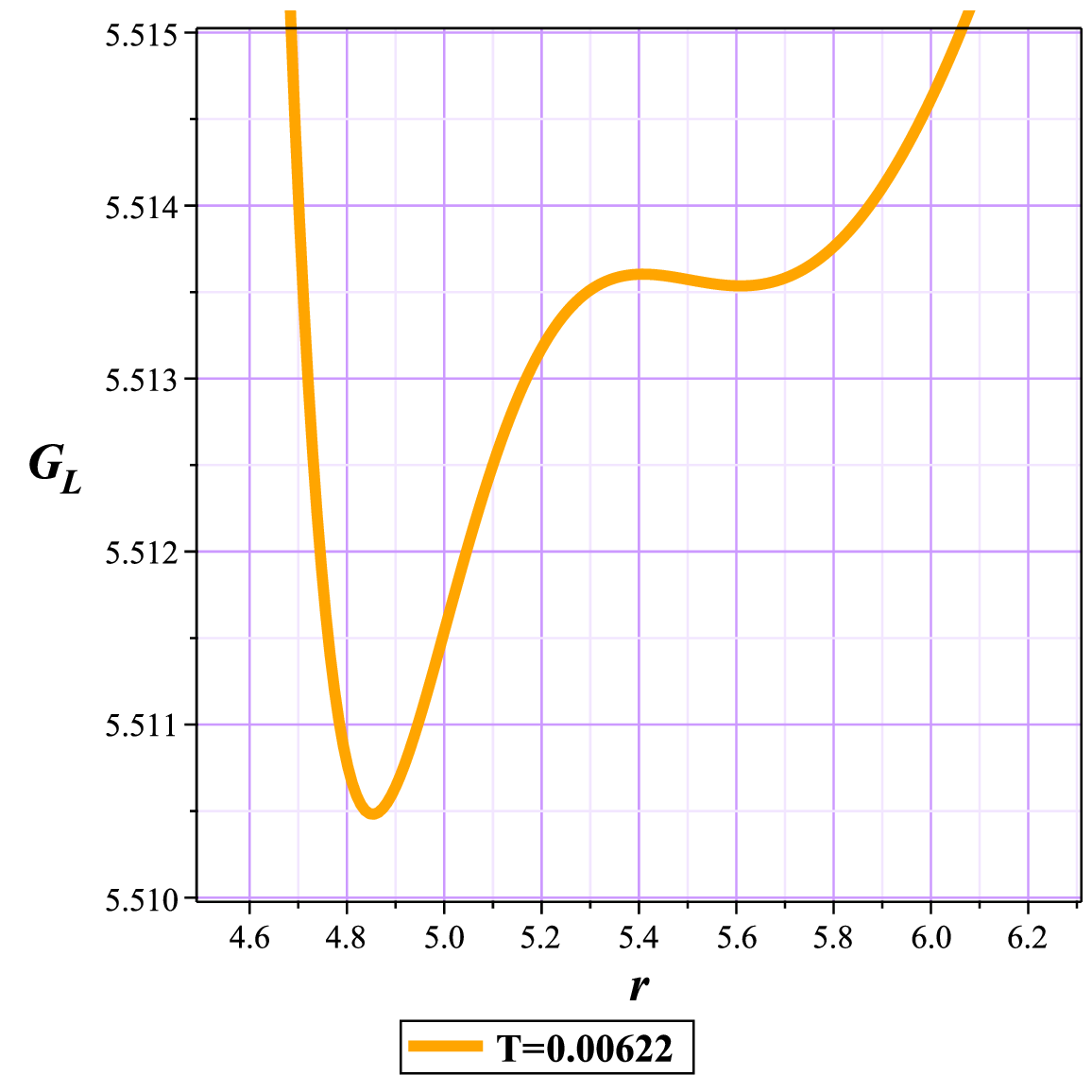}
 \label{3a}}
 \subfigure[]{
 \includegraphics[height=6cm,width=5.6cm]{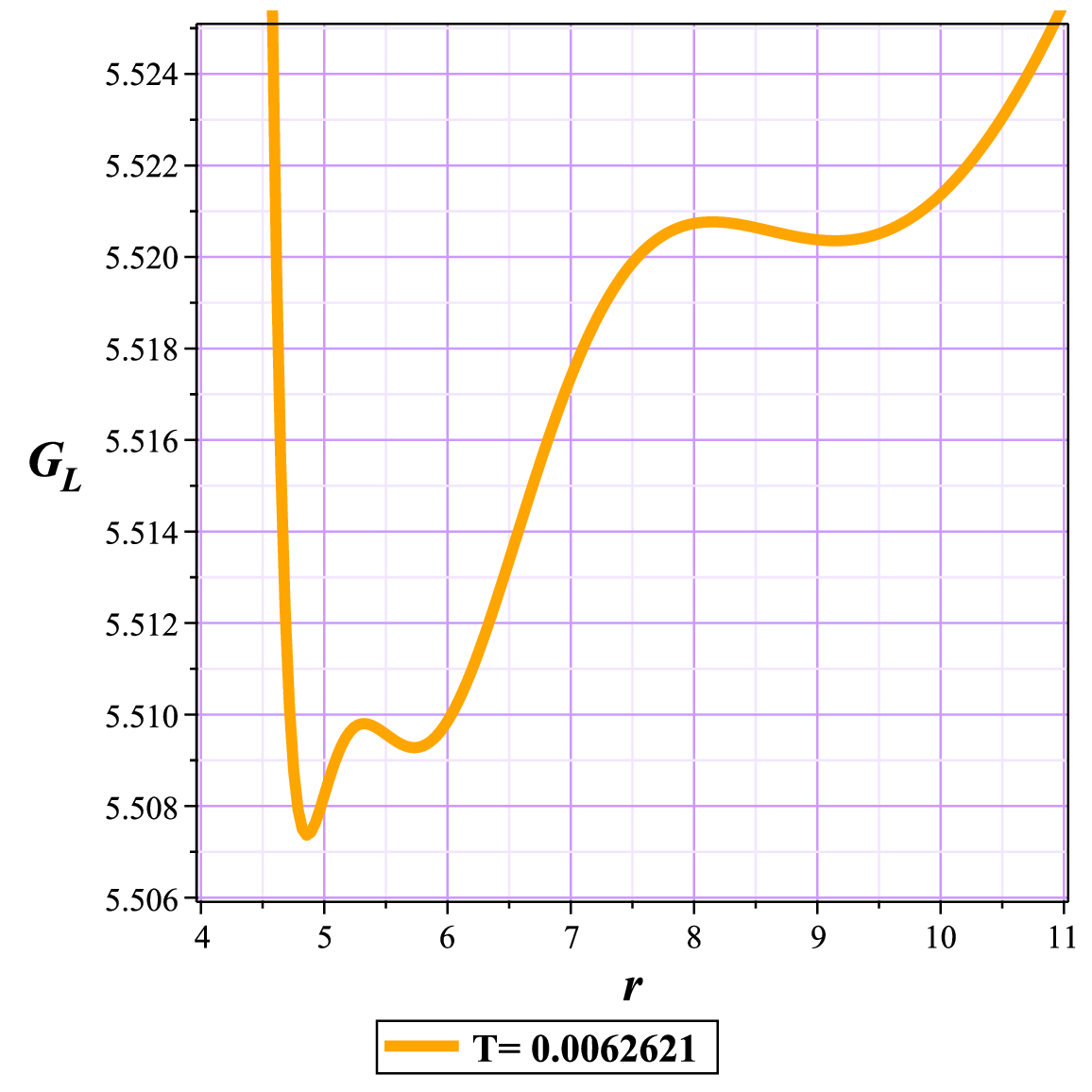}
 \label{3b}}
 \subfigure[]{
 \includegraphics[height=6cm,width=5.6cm]{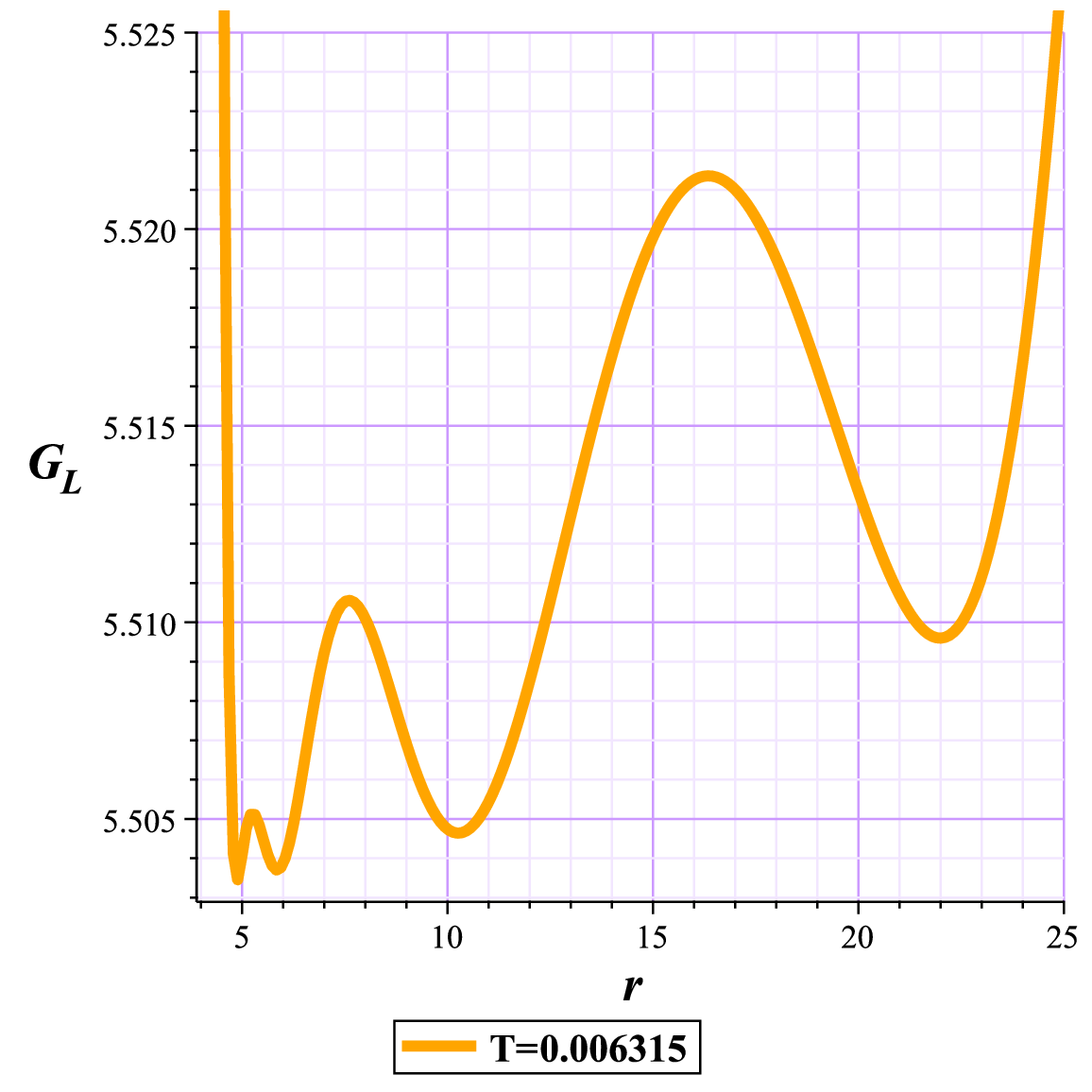}
 \label{3c}}
 \subfigure[]{
 \includegraphics[height=6cm,width=5.6cm]{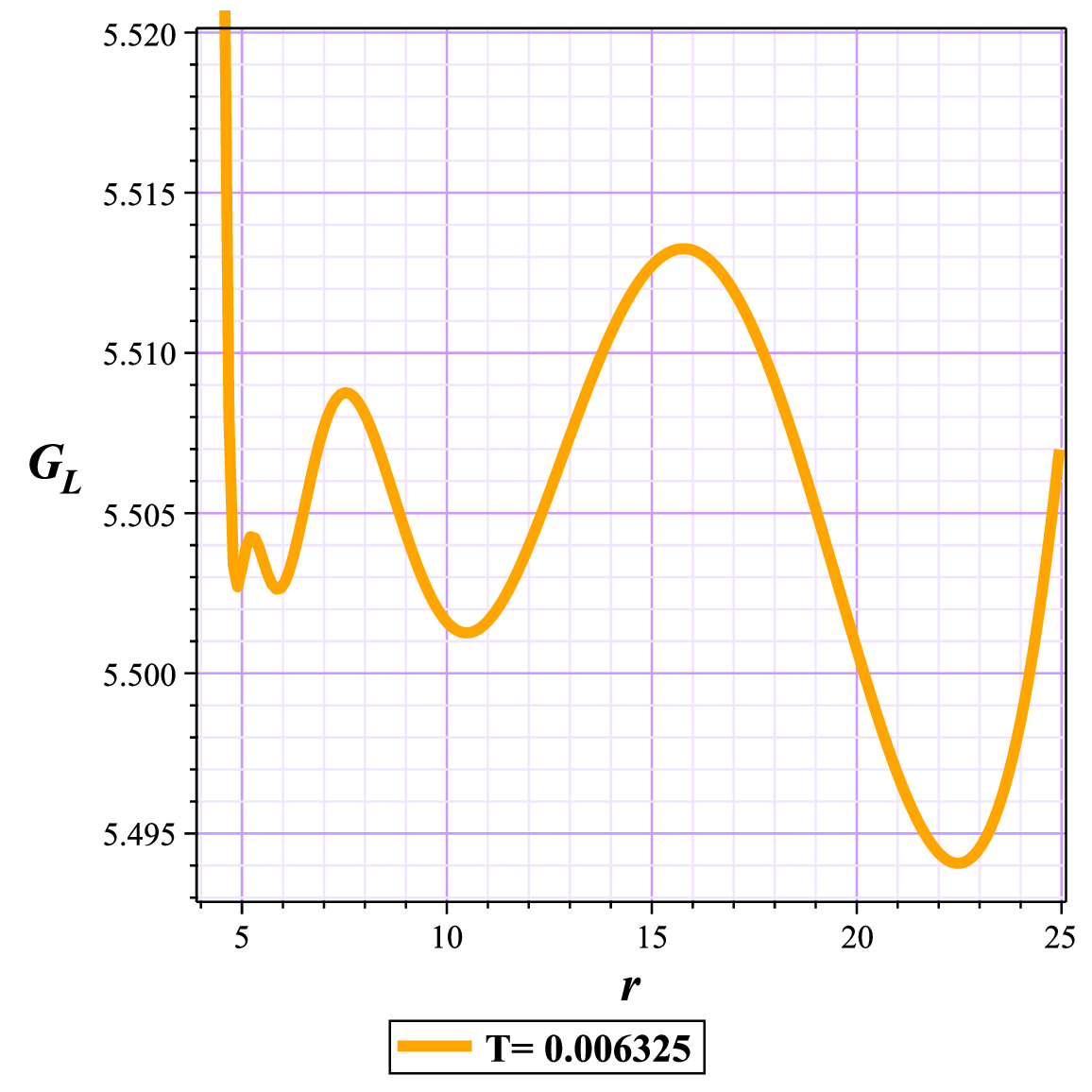}
 \label{3d}}
 \subfigure[]{
 \includegraphics[height=6cm,width=5.6cm]{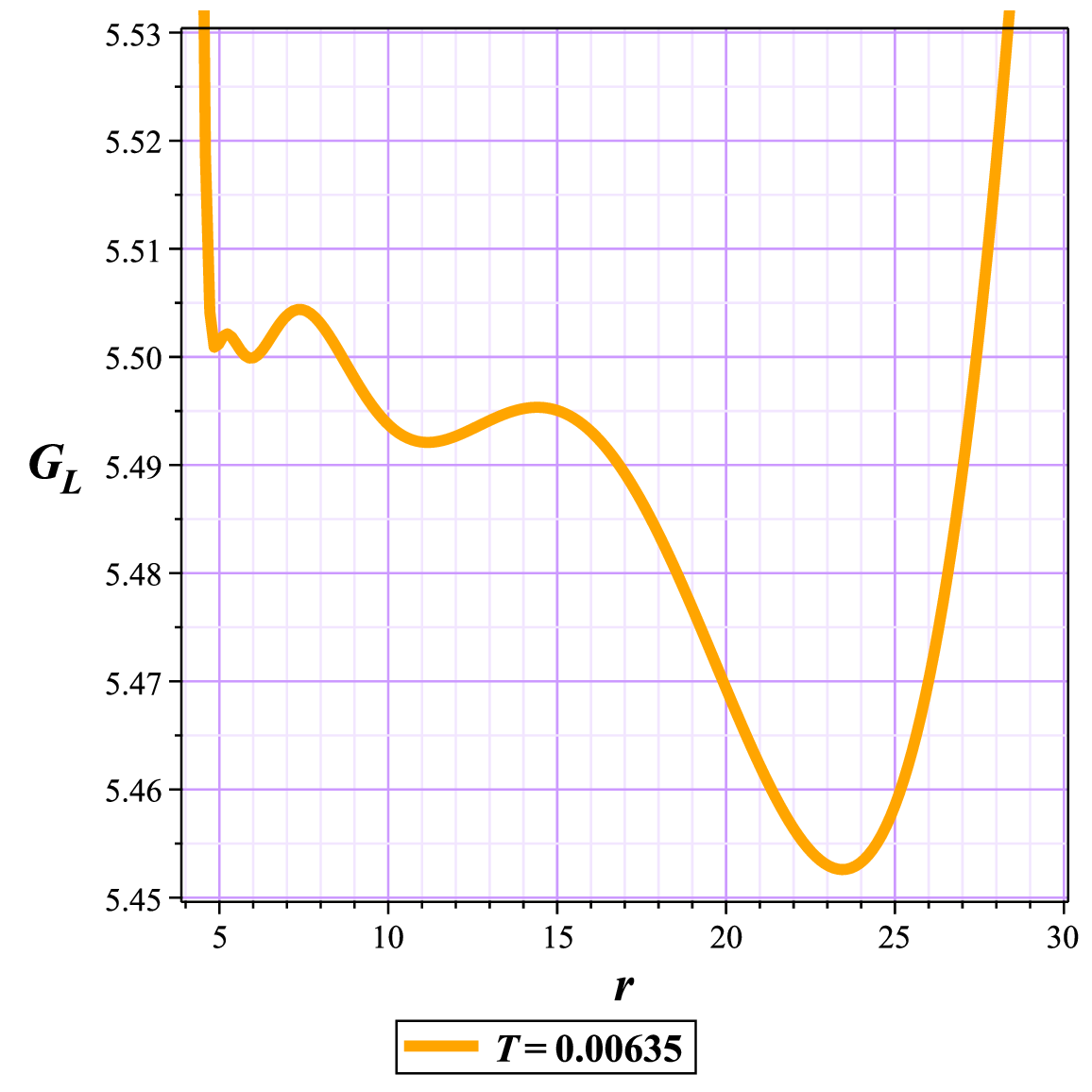}
 \label{3e}}
 \subfigure[]{
 \includegraphics[height=6cm,width=5.6cm]{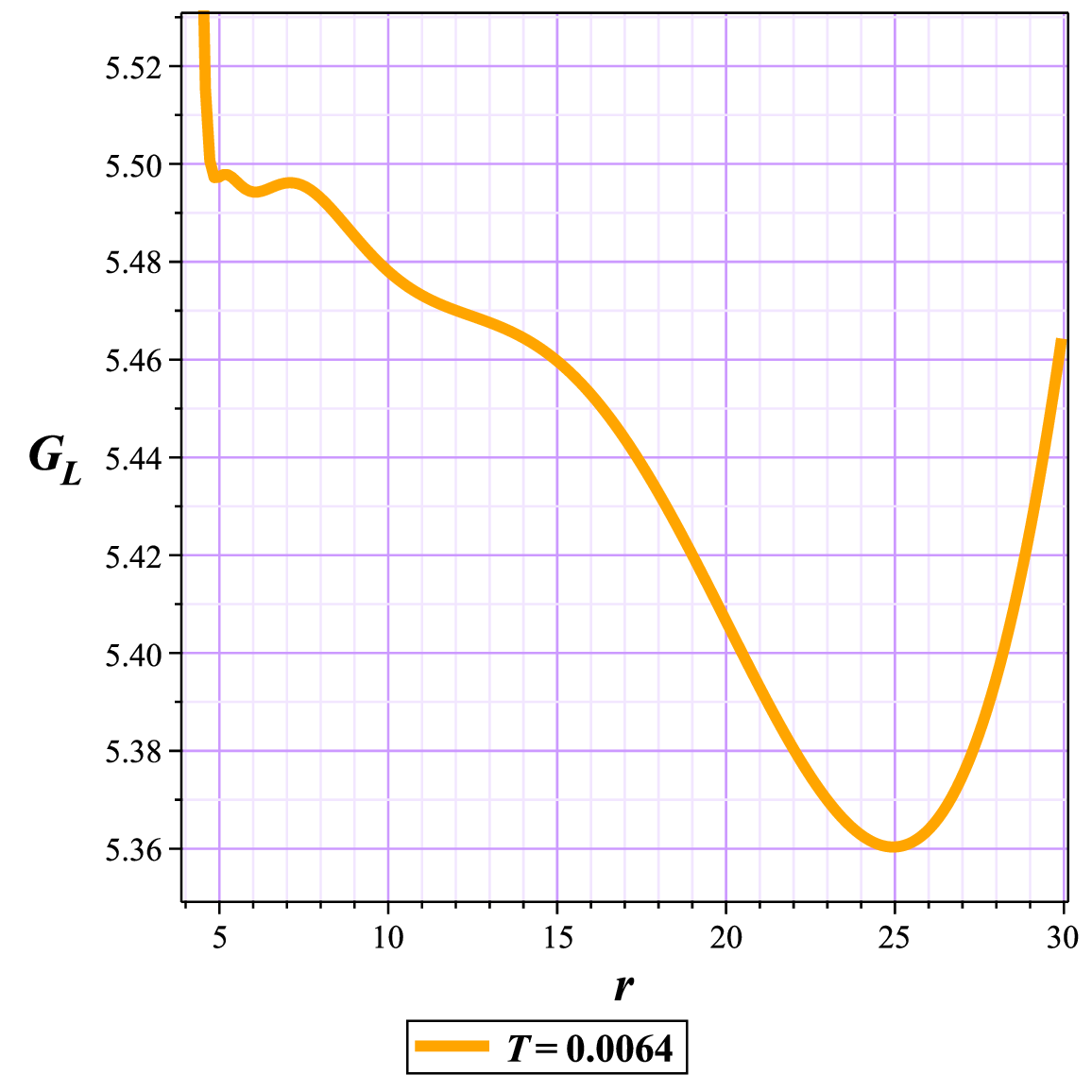}
 \label{3f}}
 \subfigure[]{
 \includegraphics[height=6cm,width=5.6cm]{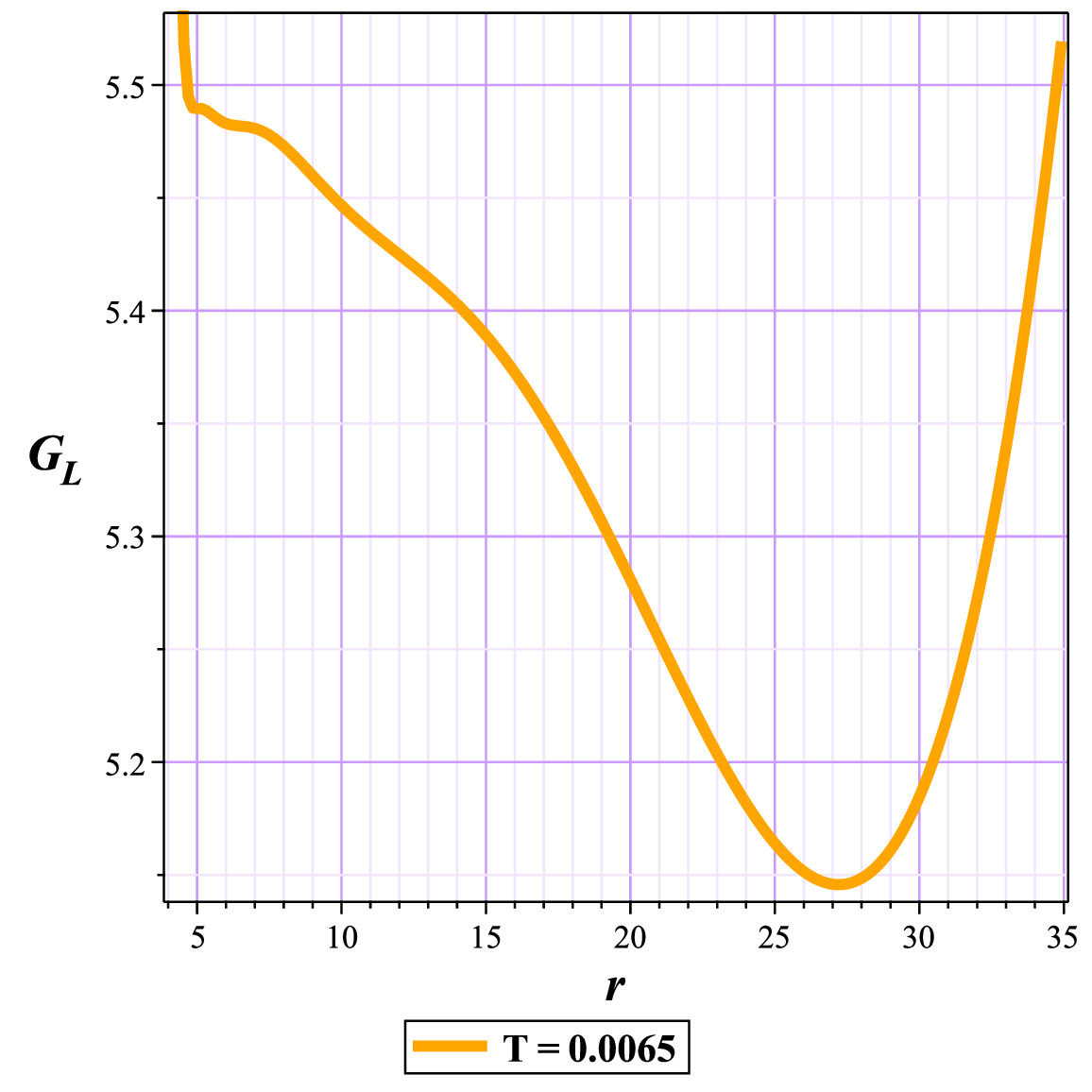}
 \label{3g}}
 \caption{\small{Behavioral sequence of energy in terms of r for different temperatures}}
 \label{m3}
\end{center}
 \end{figure}

In this section, we do not aim to analyze the phase behavior of the model by varying its parameters above or below critical points, as this has already been thoroughly studied  \cite{48,61}. Instead, based on a previously determined critical point and specific fixed parameter values, we will directly investigate the first-order phase transition dynamics using Kramers' escape rate. The primary objective of this study is to examine and compare the escape rate behavior within the first-order phase transition regime.
Therefore, utilizing Eqs. (\ref{M14}) to Eqs. (\ref{M19}), we define the following parameters for our phase transition analysis
\begin{center}
\shadowbox{\parbox{0.75\textwidth}{
$\alpha_{2} = -20.63286635$, $ \alpha_{3} = 1379.056050$, $\alpha_{4} = -133263.0329$, $\alpha_{5} = 1.550137197\times10^7$ $\alpha_{6} = -2.017480713\times10^9$, $\alpha_{7} = 5.046133724\times10^{21}$}}
\end{center}  
\newpage
with respect to these $\alpha$ s we have
\begin{center}
\shadowbox{\parbox{0.75\textwidth}{
$b_{5} = -5078.980608$, $b_{9} = 6.05480483\times10^6$, $b_{13} = -4.1315099\times10^9$, $b_{17} = 1.399821\times10^{12}$ $b_{21} = -2.0162\times10^{14}$, $b_{25} = 1.02\times10^{16}$}}
\end{center} 
In this study, rather than plotting G(T) and analyzing the phase structure through the swallowtail diagram, we will utilize the temperature diagram directly for phase transition analysis.To achieve this, considering $P=6.9\times10^{-5}$ and q=6.7511, for T diagram we have:
\begin{figure}[H]
 \begin{center}
 \includegraphics[height=8.5cm,width=14.5cm]{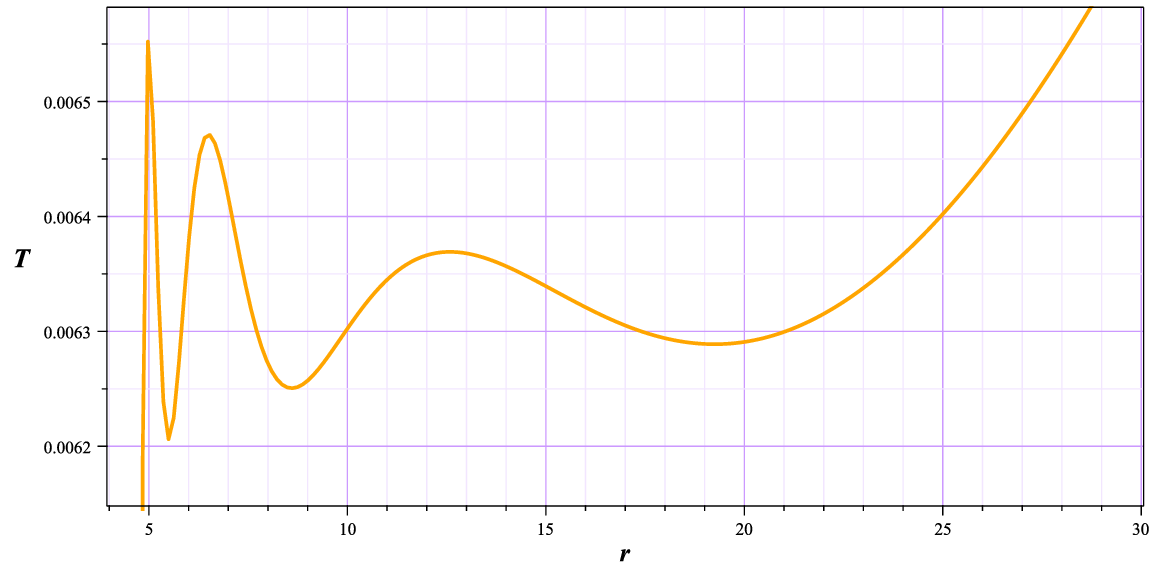}
 \caption{\small{The T diagram for 4D-Einstein-nonlinear electrodynamics Black hole }}
 \label{m2}
\end{center}
\end{figure}
As clearly illustrated in Fig. (\ref{m2}), the system exhibits diverse thermal behavior at this fixed pressure, allowing for a distinct identification of both local minima (stable black holes) and local maxima (unstable black holes). These extrema serve as critical markers for analyzing phase transition dynamics. In this study, we will classify stable black holes in ascending order of size as "smallest", "small", "large", and "largest", while unstable black holes will be assigned  numerical labels for clarity in further analysis. Given the specified parametric values, it is insightful to examine the evolution of the Gibbs free energy landscape during a first-order phase transition along the radial direction.
\begin{equation}\label{(M31)}
\begin{split}
&G_{L}=\frac{1120 P \pi  r^{28}+420 q^{2} r^{24}+420 r^{26}+210 q \,r^{20} b_{5}+140 q \,r^{16} b_{9}+105 q \,r^{12} b_{13}+84 q \,r^{8} b_{17}+70 q \,r^{4} b_{21}+60 q b_{25}}{840 r^{25}}\\ 
&-r^{2} \pi T
\end{split}
\end{equation}
This frame-by-frame perspective(Fig. (\ref{m3})), in the overall view and the increase in radius in each single frame, corresponds to an entropy increase, as indicated by Equation Eqs. (\ref{M28}). Since the directional arrow of time and entropy coincide, this progression inherently reflects the temporal evolution of the system throughout the phase transition.
\subsection{T = 0.00622}
\begin{figure}[H]
 \begin{center}
 \subfigure[]{
 \includegraphics[height=6.5cm,width=14.5cm]{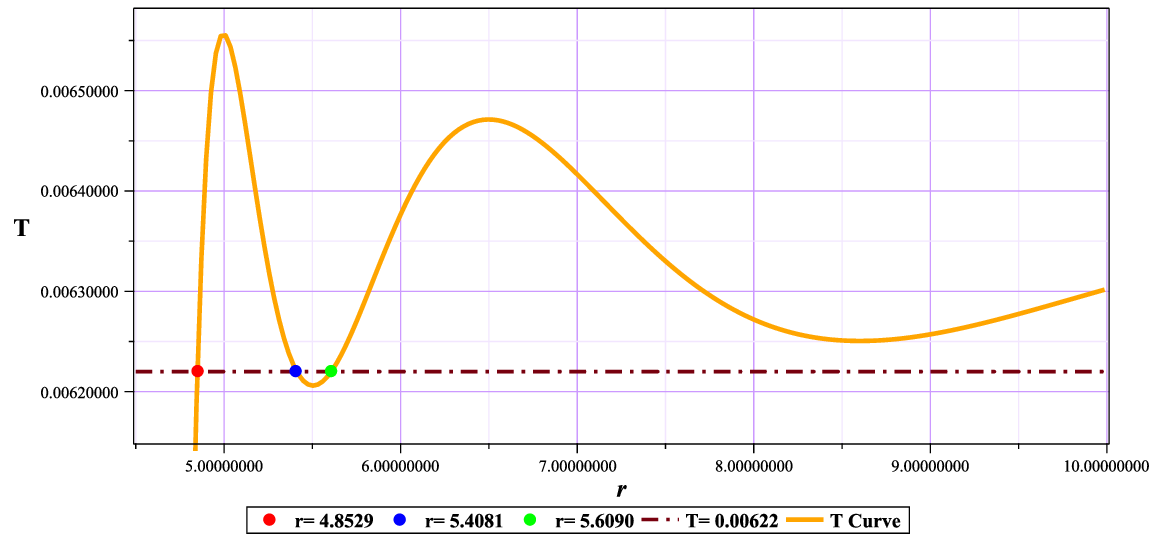}
 \label{4a}}
 \subfigure[]{
 \includegraphics[height=6.5cm,width=7cm]{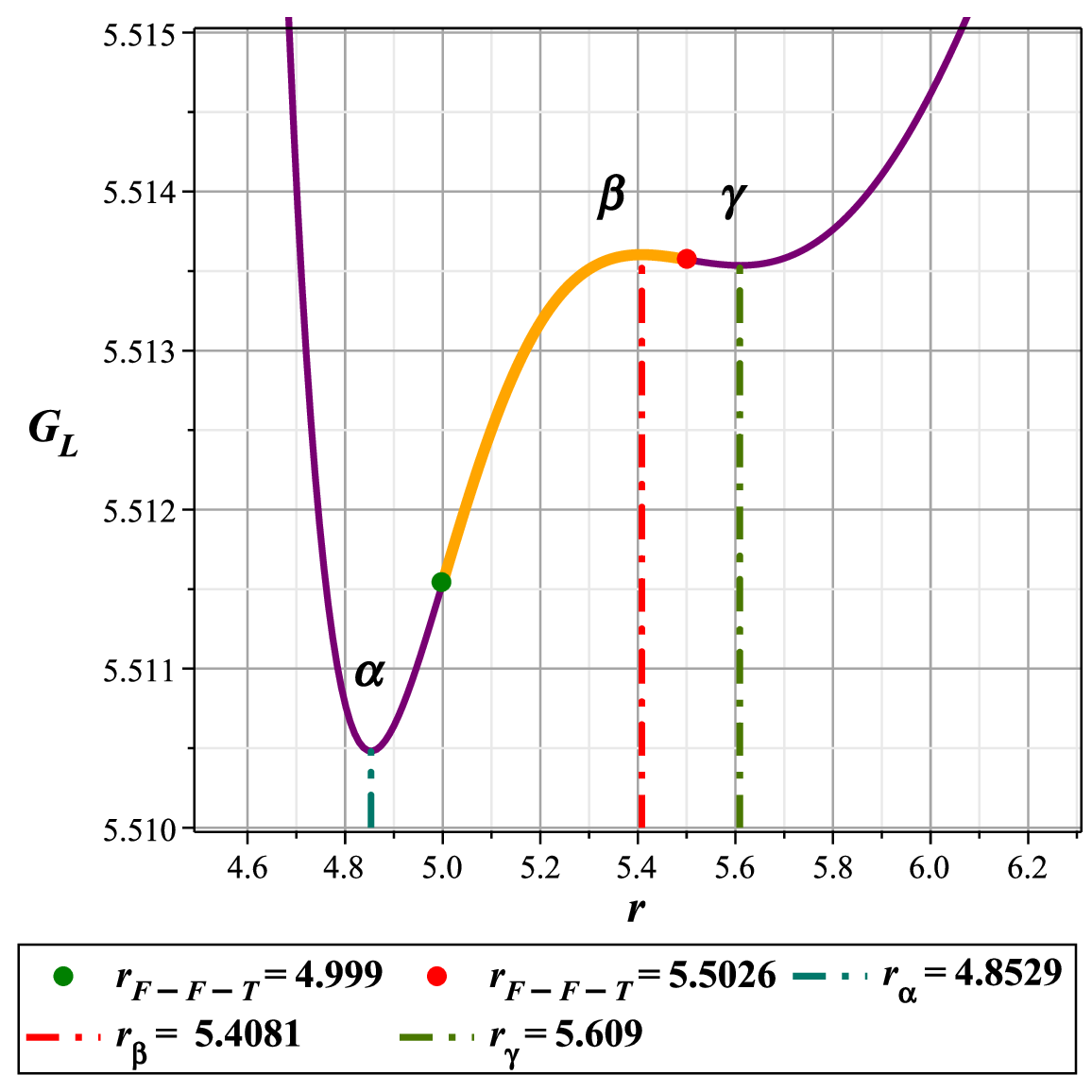}
 \label{4b}}
 \subfigure[]{
 \includegraphics[height=6.5cm,width=7cm]{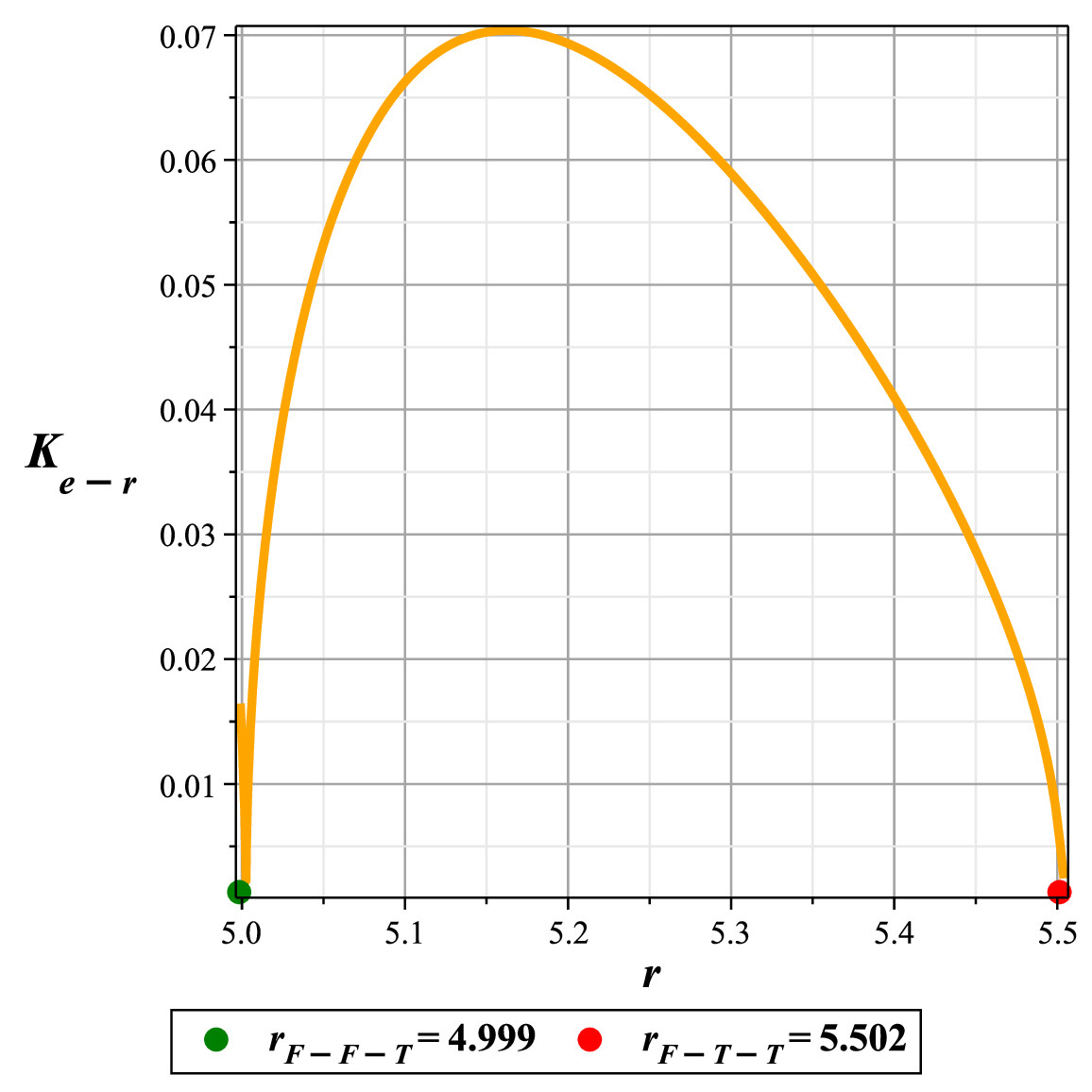}
 \label{4c}}
 \subfigure[]{
 \includegraphics[height=6.5cm,width=7cm]{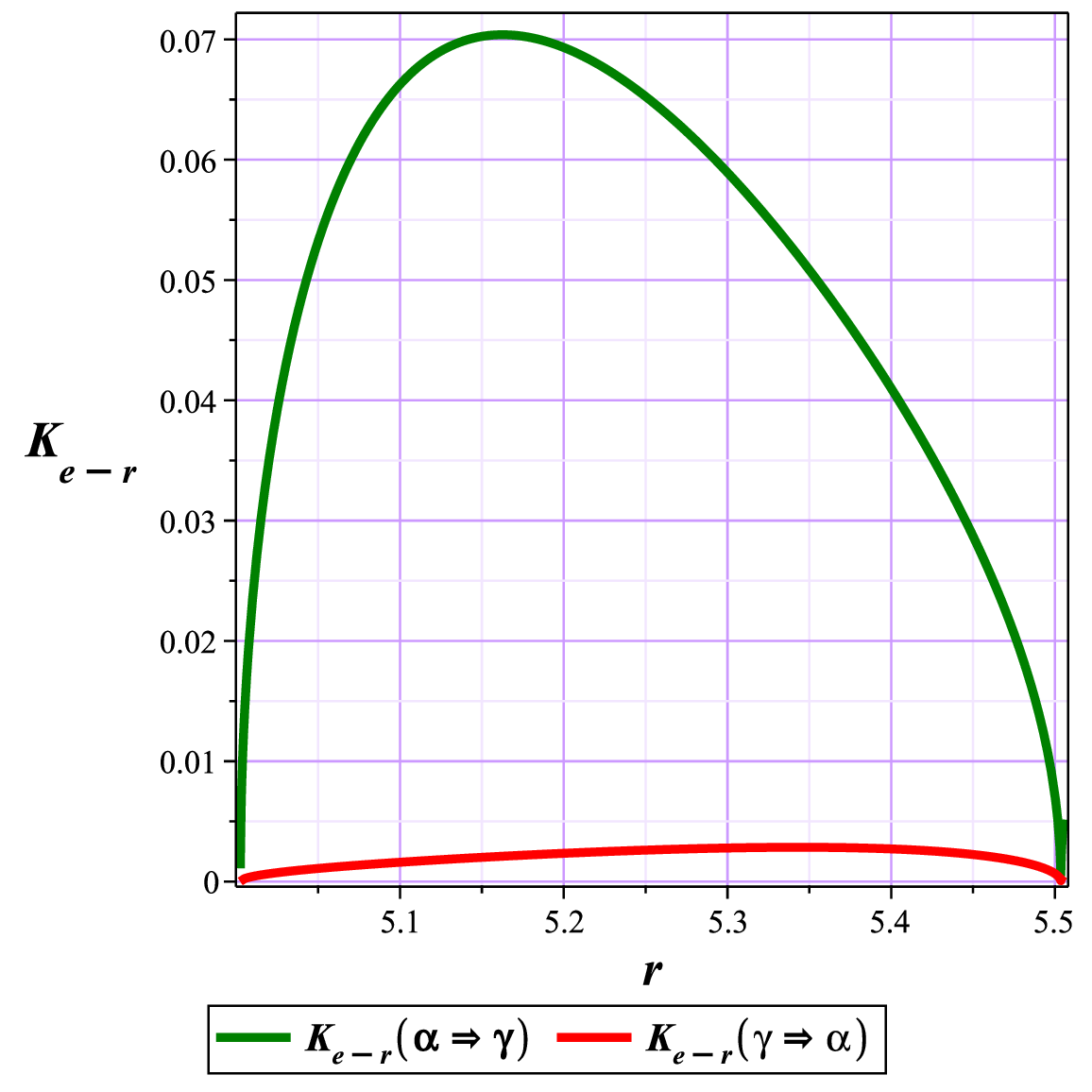}
 \label{4d}}
   \caption{\small{4a: The T diagram ,\hspace{0.1cm}4b: The graph of $G_{L}$ against r for the free chosen  parameters and the FPT points, \hspace{0.1cm}4c: Kramers escape rate during the FPT with respect to $r$ ,\hspace{0.1cm} 4d: comparing $K_{e-r}$ with respect to r for$(\alpha \rightsquigarrow \gamma)$ and $(\gamma \rightsquigarrow \alpha)$.}}
 \label{m4}
\end{center}
\end{figure}
As illustrated in Fig. \ref{m3}, the system undergoes a sequential transformation during the initial stages of the phase transition:
*In the early moments (Fig. \ref{3a}), only the "smallest" and "small" black holes have formed, while the intermediate black hole-1 is clearly discernible between the two in the form of a local and unstable maximum.
*Over time, the "large" black hole emerges (Fig. \ref{3b}), marking the next stage of evolution.
*Eventually, the "largest" black hole appears (Fig. \ref{3c}), completing the primary sequence of black hole formation.
However, the process does not conclude with the emergence of the largest black hole. Since it has not yet evolved into a global minimum, the transition will continue until this black hole reaches a globally stable state (Fig. \ref{3g}).
Now, using $G_{L}$ as an effective potential that from one hand is considered to be a driver and, in a way, a controller of the phase transition, and on the other hand, in Kramers view, it will act as a potential well that the black hole must pass through due to thermal fluctuation, from the displayed evolution frames of  $G_{L}$, we will go to those that have a more prominent role and with respect to Kramers view analyze them carefully.
As demonstrated in Fig. \ref{4a}, the temperature curve with line T = 0.00622 intersects at three distinct points, indicating that the $G_{L}$ function possesses three extrema. Consequently, we anticipate the presence of two stable black holes (designated as smallest and small) and one unstable black hole, a prediction confirmed in Fig. \ref{4b}.  Upon plotting the escape rate within the phase transition regime (Fig. \ref{4c}), we observe a striking alignment between the escape rate and the beginning and end of the transition process. Specifically, the escape rate increases from its minimum value at the phase transition onset and subsequently declines to its minimum again at the transition’s conclusion.  
This remarkable consistency is highly significant, as it demonstrates that even when adopting a particle-based perspective—using escape rate dynamics to study black hole phase behavior—the resulting equations precisely guide and regulate system transformations within the same phase transition regime predicted by classical studies. But is there any other useful information that can be extracted from the escape rate diagram?
While Fig. \ref{4c} provides a clear representation of the escape rate within the phase transition regime, a closer inspection of Fig. \ref{4b} reveals an important nuance—the starting point of the phase transition is slightly elevated compared to the nearby local minimum ($\alpha$). This naturally raises an intriguing question: Since the escape rate starts from a minimum at the starting point of the phase transition, how does the escape process from $\alpha$ to $\gamma$ take place?
\begin{figure}[H]
 \begin{center}
 \includegraphics[height=7.5cm,width=14.5cm]{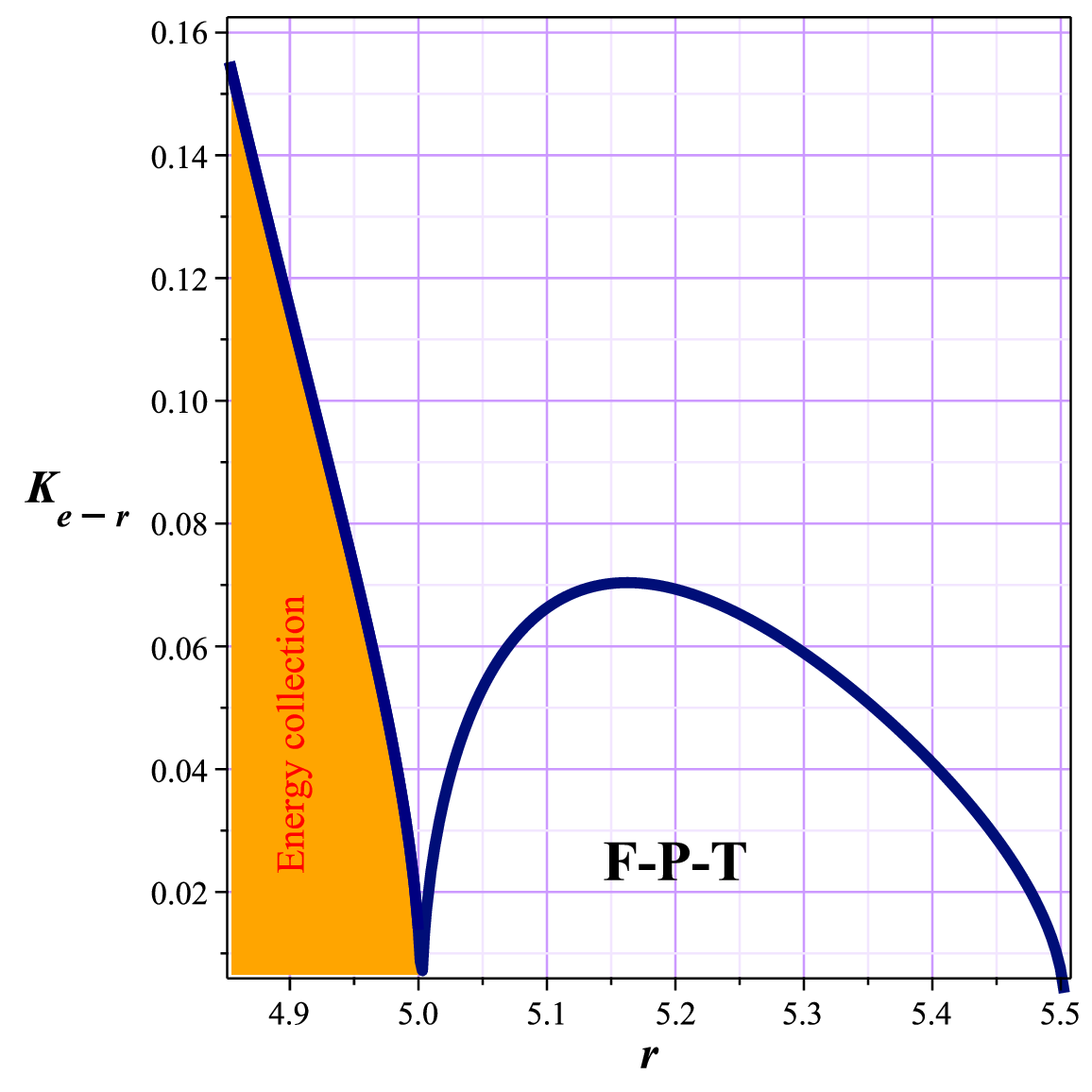}
 \caption{\small{Kramers escape rate during $\alpha\rightsquigarrow\gamma$ transition with respect to $r$ }}
 \label{m5}
\end{center}
\end{figure}
To answer should state that, escape probability  fundamentally depends on whether thermal fluctuations provide sufficient energy for the system to overcome potential barriers in a single direct leap. As observed in Fig. \ref{m5}, the energy landscape appears insufficient to support a direct escape , suggesting an alternative interpretation of the system’s behavior. In this scenario, the system can be modeled as a stretched spring, implying that the black hole, during its evolutionary process, first undergoes radial expansion (corresponding to an increase in temperature), effectively accumulating energy up to a tolerable threshold. Once the system reaches this critical tolerance limit, a phase transition occurs. The radial tolerable range of each black hole—with respect to the nearest local minimum—can be inferred not only by analyzing the Gibbs free energy diagram also by using the escape rate diagram and equations. 
Specifically:  If the system exhibits an incomplete or complete semicircle in escape rate graph (such as the orange-shaded region in Fig. \ref{m5} or a curve resembling Fig. \ref{4c}) yet lacks any extremum in the energy function, then this region represents an energy accumulation state without an actual phase transition.  Conversely, if the energy function contains the extremums within the studied range and the escape rate forms a complete circle (minimum rate to minimum rate), then this strongly predicts the occurrence of a first-order phase transition. These insights provide valuable information regarding the mechanisms governing black hole phase transitions, reinforcing the connection between thermal fluctuations, escape dynamics, and potential energy landscapes.
\subsection{T = 0.0062621}
\begin{figure}[H]
 \begin{center}
 \subfigure[]{
 \includegraphics[height=8.5cm,width=8.5cm]{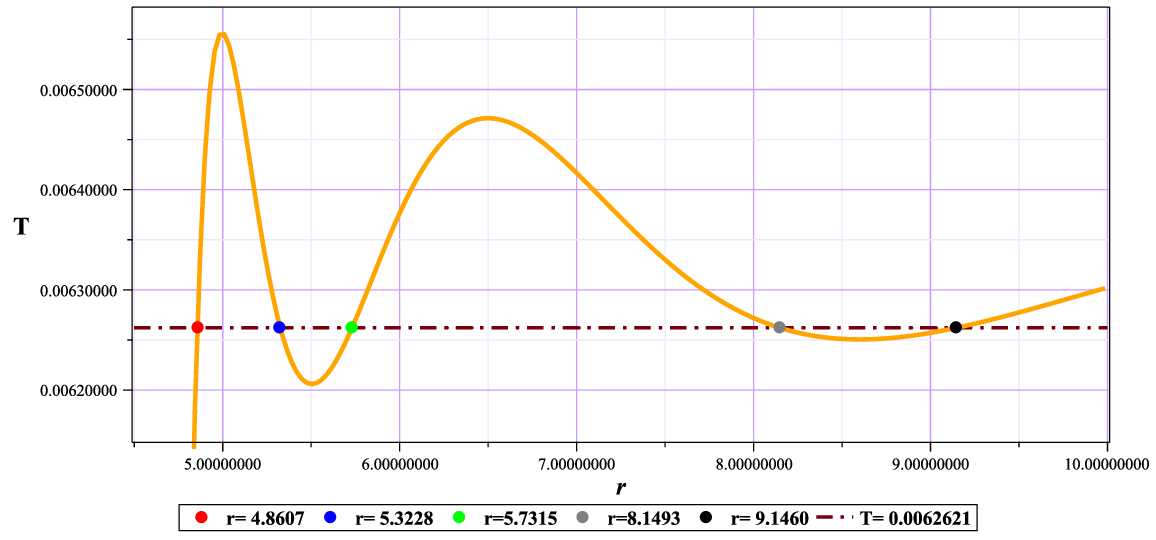}
 \label{6a}}
 \subfigure[]{
 \includegraphics[height=8.5cm,width=8.5cm]{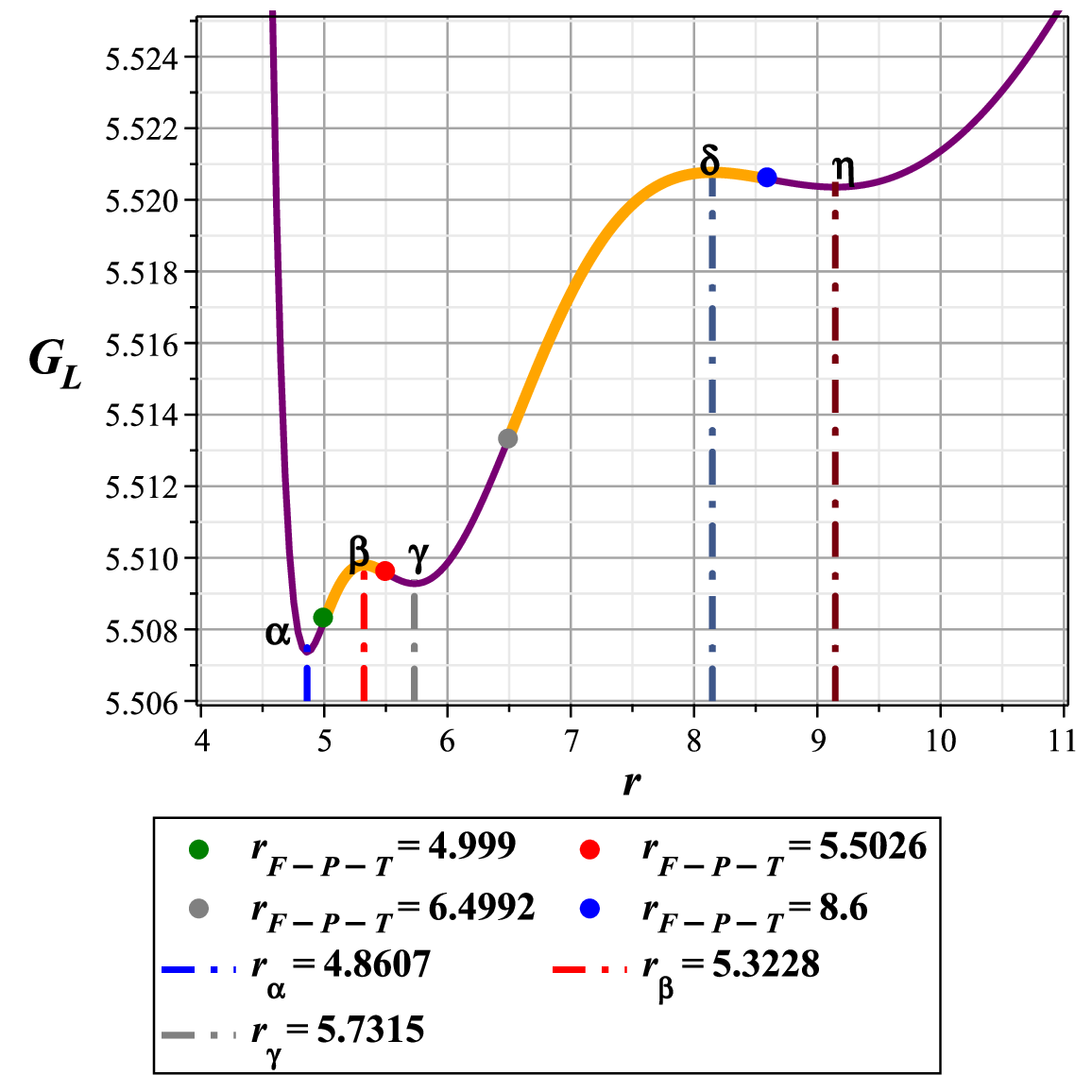}
 \label{6b}}
 \subfigure[]{
 \includegraphics[height=8.5cm,width=7cm]{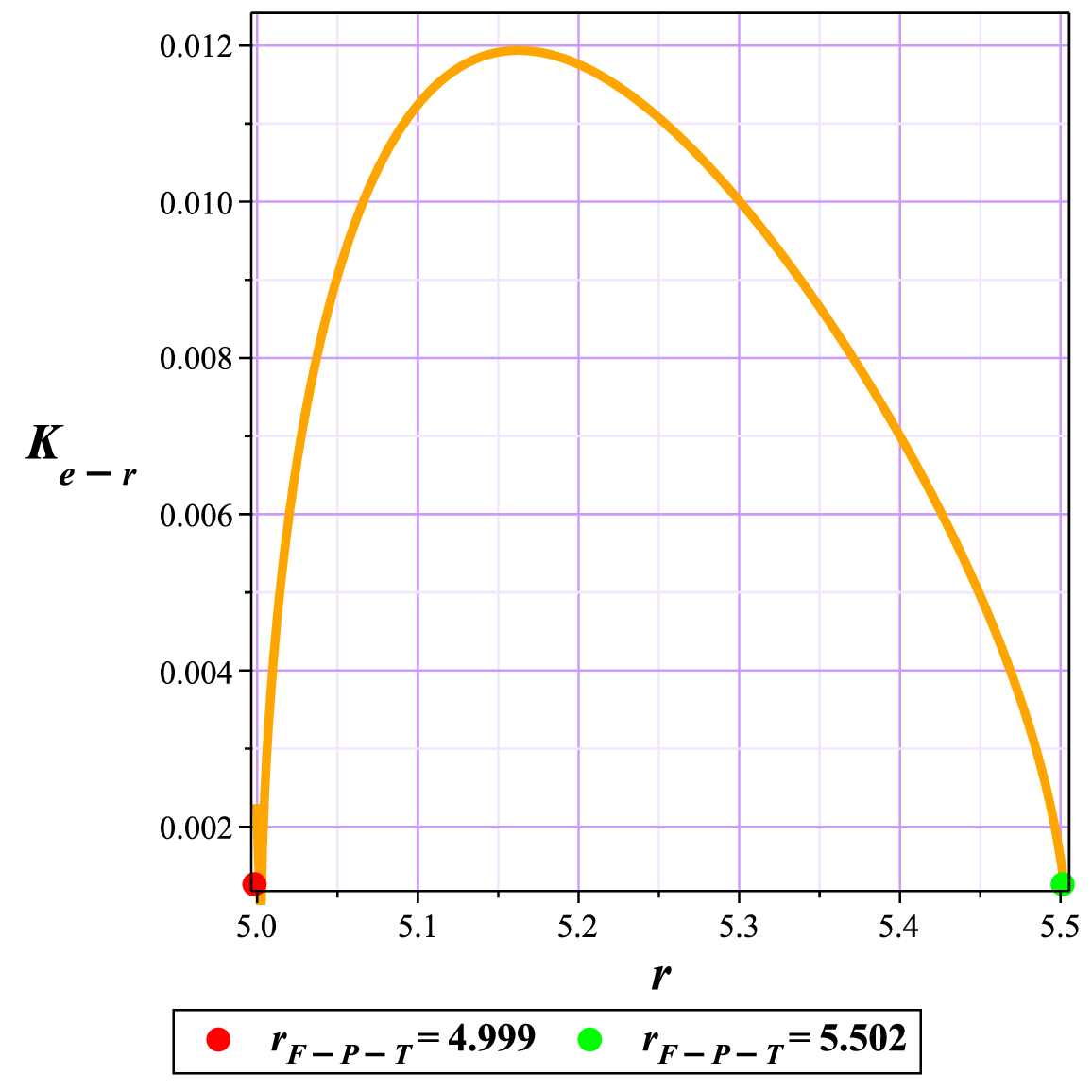}
 \label{6c}}
 \subfigure[]{
 \includegraphics[height=8.5cm,width=7cm]{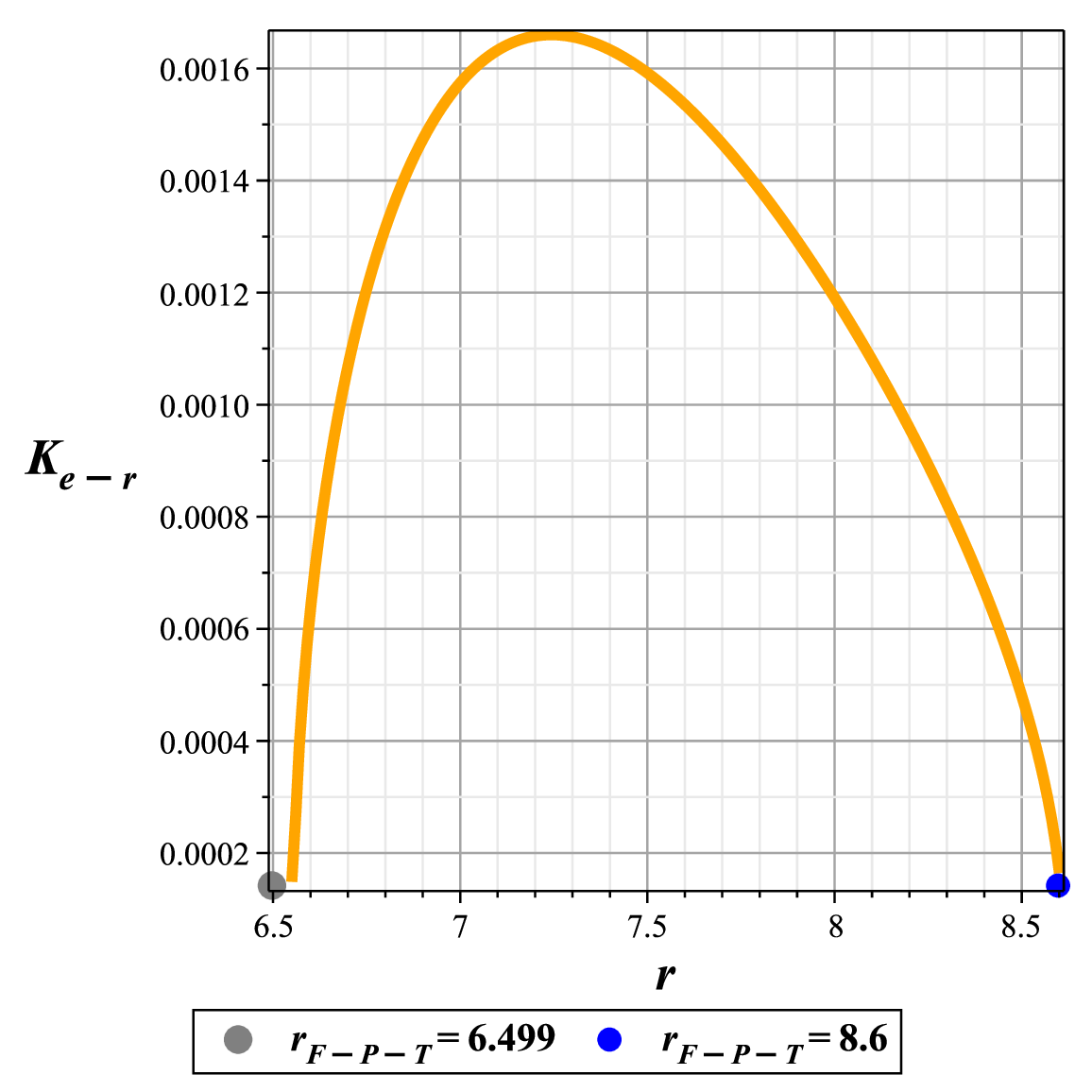}
 \label{6d}}
   \caption{\small{6a: The T diagram ,\hspace{0.1cm}6b) The graph of $G_{L}$ against r for the free chosen  parameters and the FPT points, \hspace{0.1cm}6c, 6d: Kramers escape rate during the FPT with respect to $r$.}}
 \label{m6}
\end{center}
\end{figure}
In Fig. \ref{6a}, the temperature curve with line T = 0.0062621 intersects at five distinct points, indicating that the $G_{L}$ function possesses five extrema. Consequently, we anticipate the presence of three stable black holes (designated as smallest and small, big) and two unstable black holes, a prediction confirmed in Fig. \ref{6b}.\\  Upon plotting the escape rates within the two phase transitions regime (Fig. \ref{6c},Fig. \ref{6d}), again we observe the striking alignment between these escape rates and the beginning and end of the transition processes.\\
\begin{figure}[H]
 \begin{center}
 \subfigure[]{
 \includegraphics[height=8.5cm,width=8.5cm]{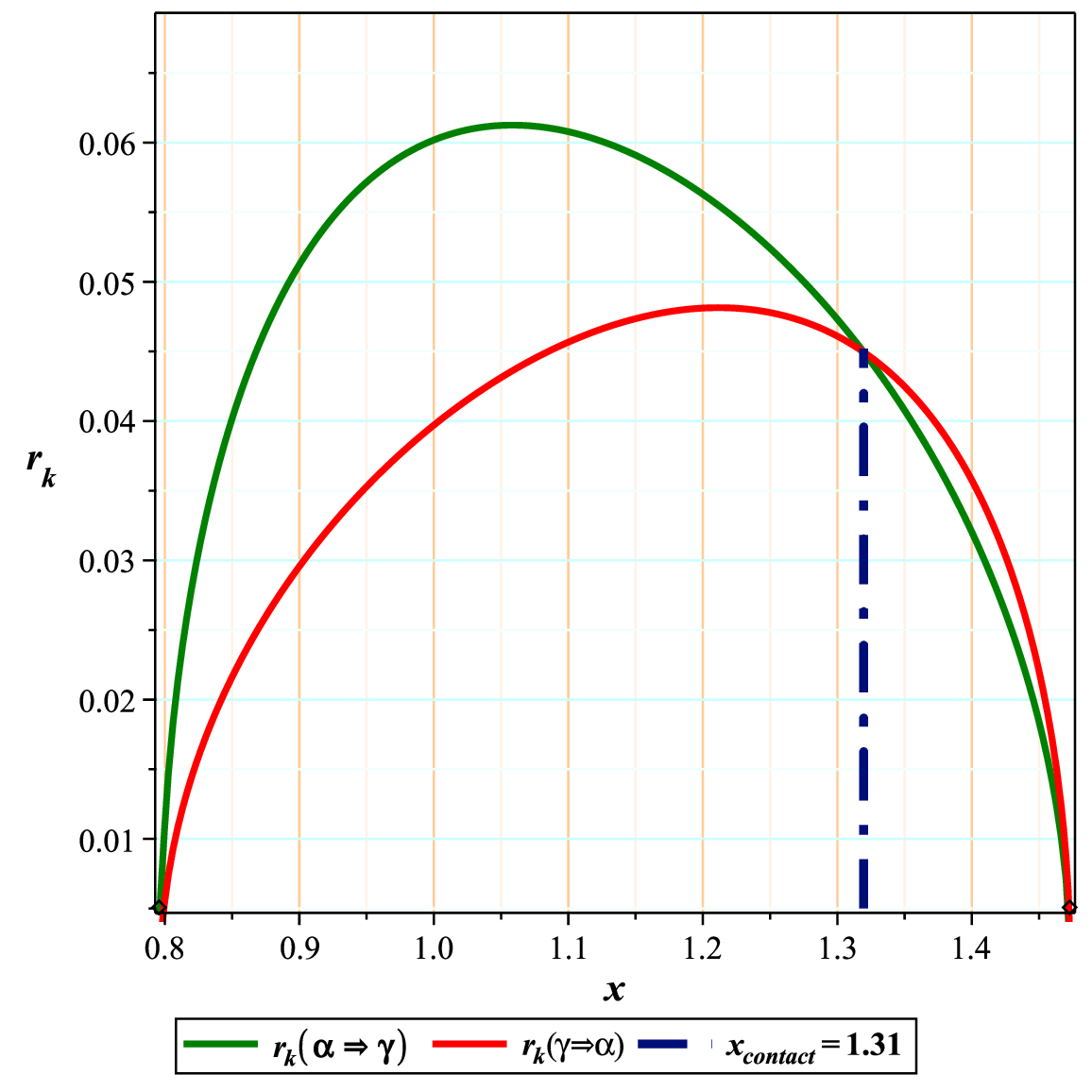}
 \label{7a}}
 \subfigure[]{
 \includegraphics[height=8.5cm,width=8.5cm]{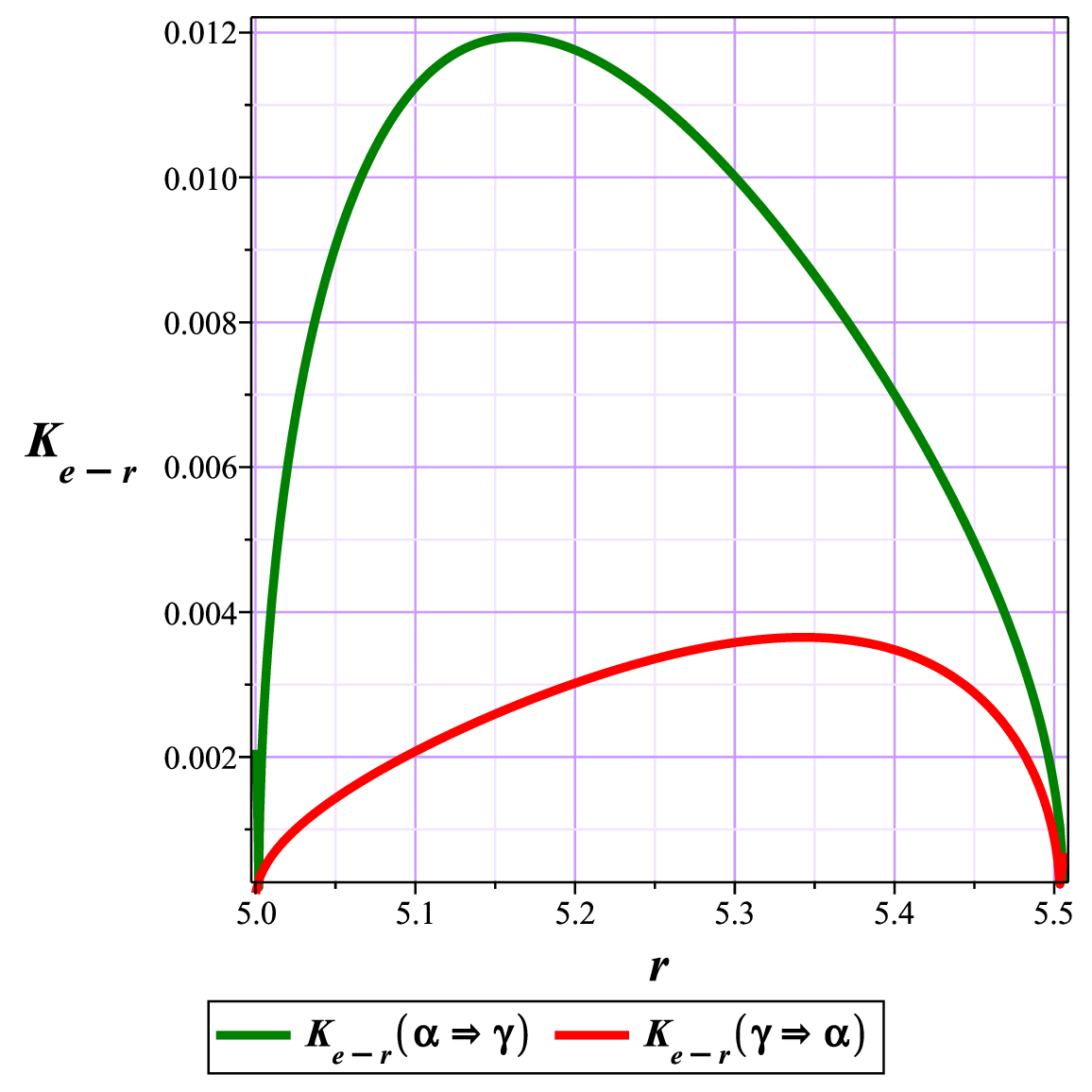}
 \label{7b}}
 \subfigure[]{
 \includegraphics[height=8.5cm,width=8.5cm]{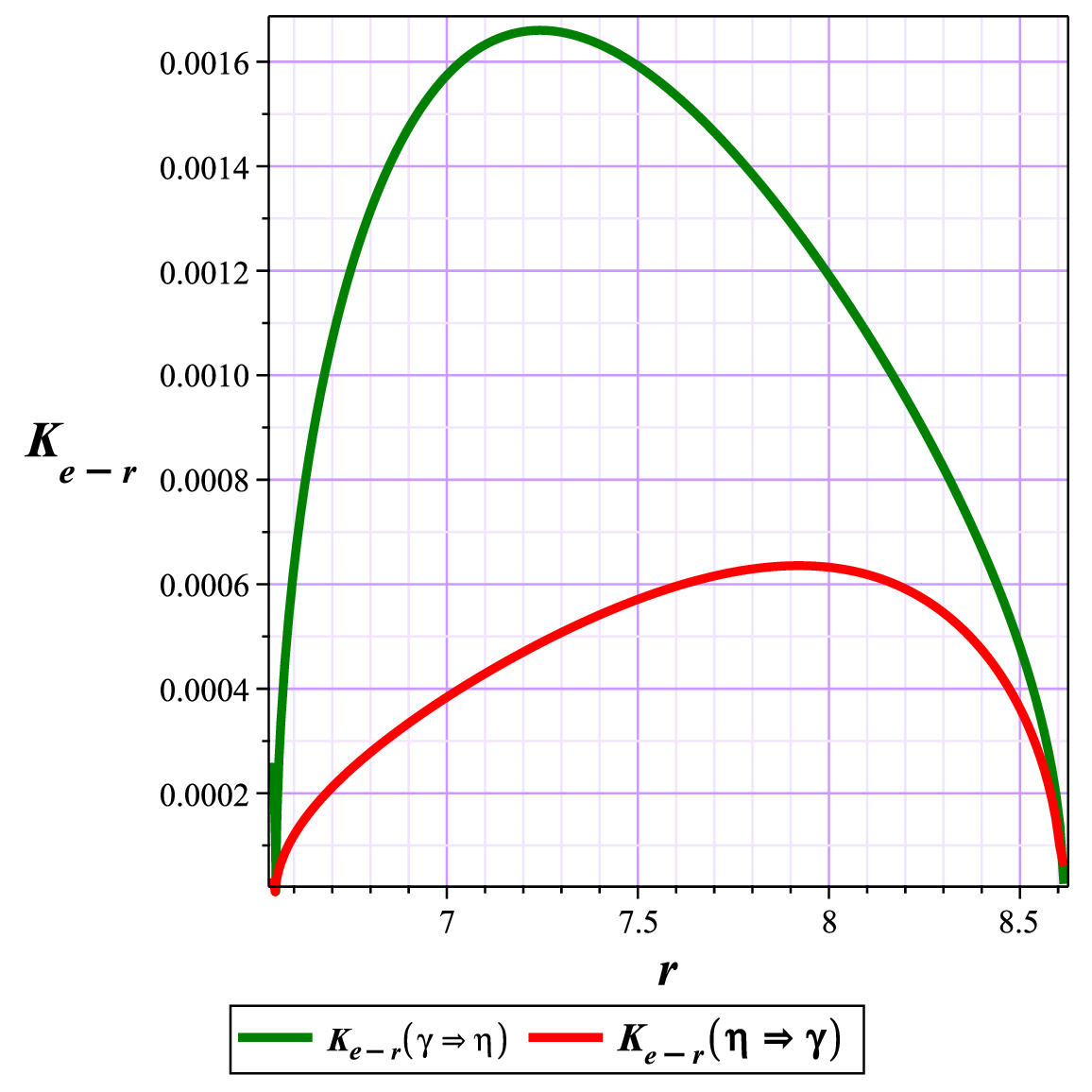}
 \label{7c}}
   \caption{\small{7a: comparing $K_{e-r}$ with respect to r for$(\alpha \rightsquigarrow \gamma)$ and $(\gamma \rightsquigarrow \alpha)$ and the coordinate of contact point for 4DAdSEinstein-gauss-bonnet-Yang-Mills black hole with a cloud of strings black hole \cite{59} ,\hspace{0.1cm}7b: comparing $K_{e-r}$ with respect to r for$(\alpha \rightsquigarrow \gamma)$ and $(\gamma \rightsquigarrow \alpha)$, \hspace{0.1cm}7c: comparing $K_{e-r}$ with respect to r for$(\gamma \rightsquigarrow \eta)$ and $(\eta \rightsquigarrow \gamma)$}}
 \label{m7}
\end{center}
\end{figure}
An intriguing and significant aspect of this study, compared to conventional first-order phase transition models, lies in the behavior observed in single-transition frameworks. In such cases, near the completion of the phase transition, a brief and localized reversal occurs, wherein the system momentarily shifts from a large black hole to a smaller one ( Fig. \ref{7a})\cite{55,56,59,60}.\\ As previously discussed, this process could potentially act as a regulatory mechanism, ensuring the orderly termination of the transition and preventing an uncontrolled continuation.\\
However, in this scenario, following the formation of a "small" black hole opposite the "smallest" one, the newly formed black hole exhibits metastable form, preventing the process from halting at this stage. Instead, the system rapidly progresses towards the formation of larger and ultimately the largest black holes.Put more precisely,  throughout the transition process, the probability of a direct progression significantly exceeds that of the reverse process, preventing the latter from manifesting.\\Consequently, it is expected that the reversal mechanism will no longer be feasible under these conditions—a prediction that is clearly supported by computational results, as illustrated in Fig. \ref{4d}, Fig. \ref{7b} and  Fig. \ref{7c}. 
\begin{figure}[H]
 \begin{center}
 \includegraphics[height=8.5cm,width=14.5cm]{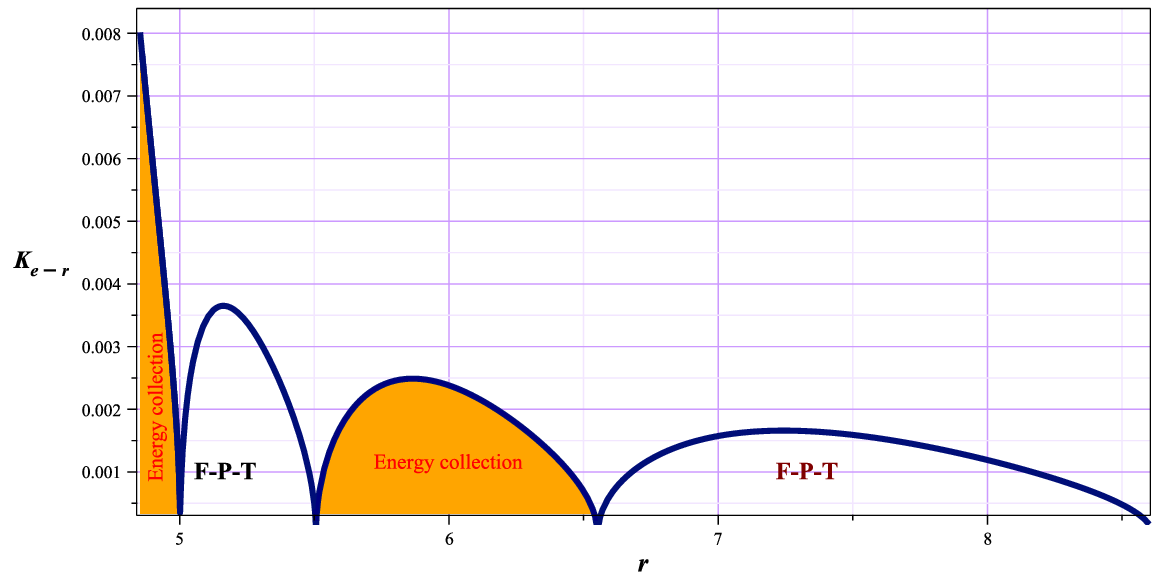}
 \caption{\small{Kramers escape rate during $\alpha\rightsquigarrow\eta$ transition with respect to $r$ }}
 \label{m8}
\end{center}
\end{figure}
As observed in Fig. \ref{m8}, the energy landscape again appears insufficient to support a direct escape transition from $\alpha$ to other local minima. Thus, the phase transition proceeds stepwise, accompanied by energy accumulation regions prior to each transition phase—highlighted in orange zones in Fig. \ref{m8}.\\
Beyond this behavior, another notable trend emerges: the progressive decrease in the maximum escape rate with gradual increases in radius at each stage (Fig. \ref{m8}). This phenomenon suggests that the system dynamically adjusts in a manner that ultimately leads the black hole to equilibrium at the global minimum, corresponding to the largest black hole configuration.
\subsection{T = 0.006325}
\begin{figure}[H]
 \begin{center}
 \subfigure[]{
 \includegraphics[height=6.5cm,width=14.5cm]{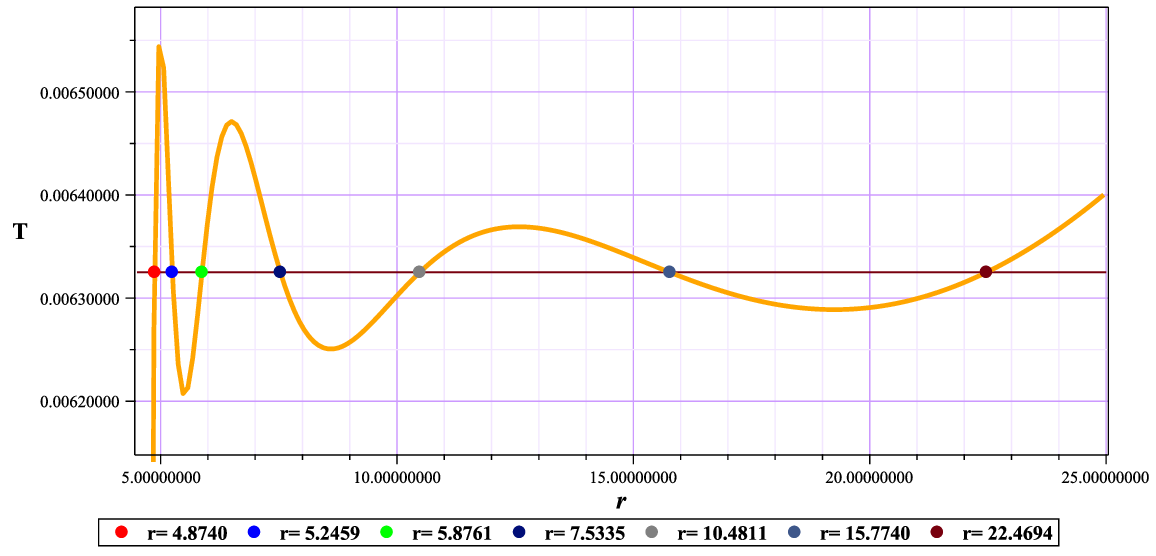}
 \label{9a}}
 \subfigure[]{
 \includegraphics[height=6.5cm,width=14.5cm]{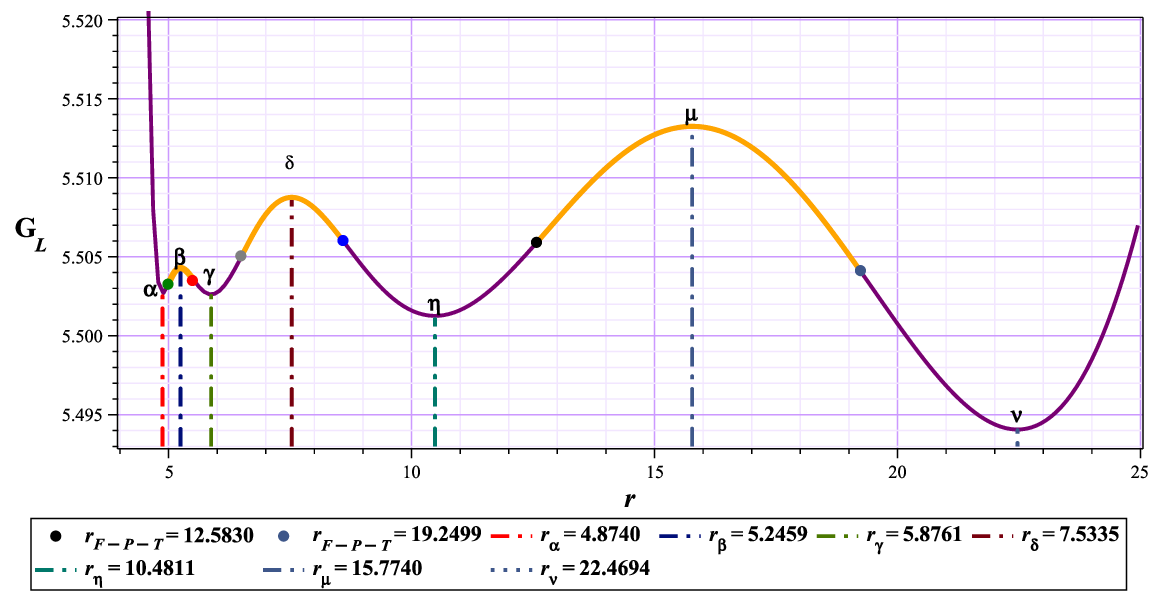}
 \label{9b}}
 \subfigure[]{
 \includegraphics[height=6.5cm,width=14.5cm]{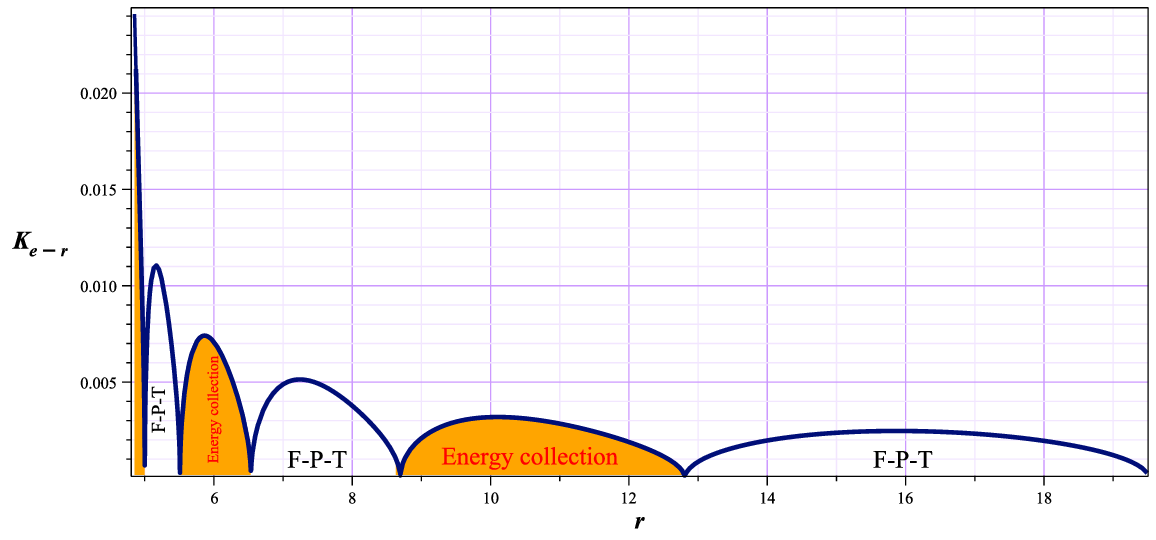}
 \label{9c}}
   \caption{\small{9a: The T diagram ,\hspace{0.1cm}9b: The graph of $G_{L}$ against r for the free chosen  parameters and the FPT points (Since the green to red and gray to blue phase transitions were shown in Figs. 4b and 6b, only the final phase transition values), \hspace{0.1cm}9c: Kramers escape rate during $\alpha\rightsquigarrow\eta$ transition with respect to r}}
 \label{m9}
\end{center}
\end{figure}

In this scenario, we encounter the highest number of extrema (Fig. \ref{m9}). Given that the graphical trends and interpretations remain consistent with previous cases—and to avoid repetition (for instance, the escape rate alignment at the beginning and end of each phase transition still holds with remarkable precision)—we limit our discussion to displaying only the most impactful graphs for analysis.
\subsection{T= 0.0064}
\begin{figure}[H]
 \begin{center}
 \subfigure[]{
 \includegraphics[height=6.5cm,width=14.5cm]{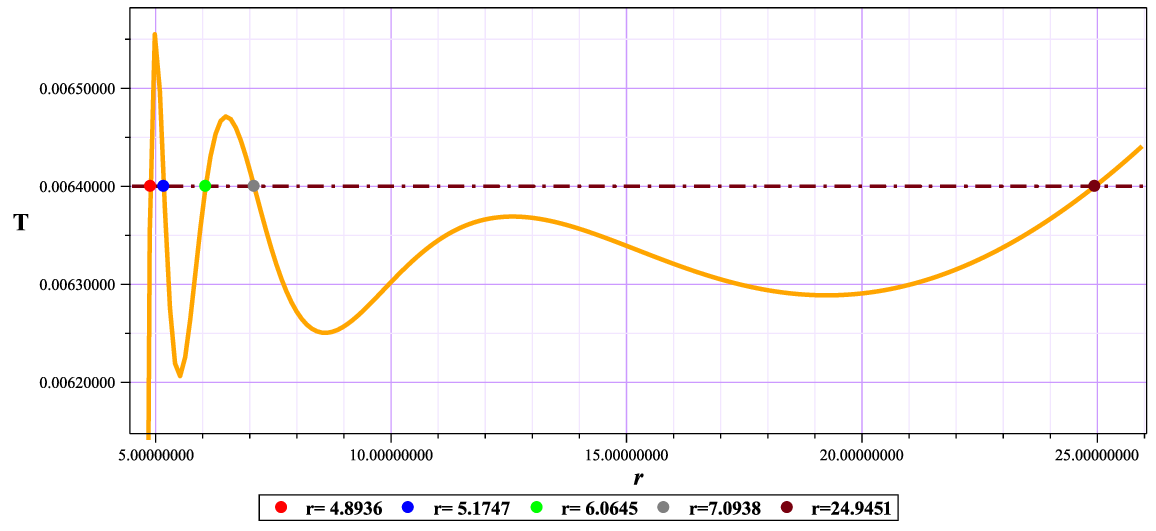}
 \label{10a}}
 \subfigure[]{
 \includegraphics[height=6.5cm,width=14.5cm]{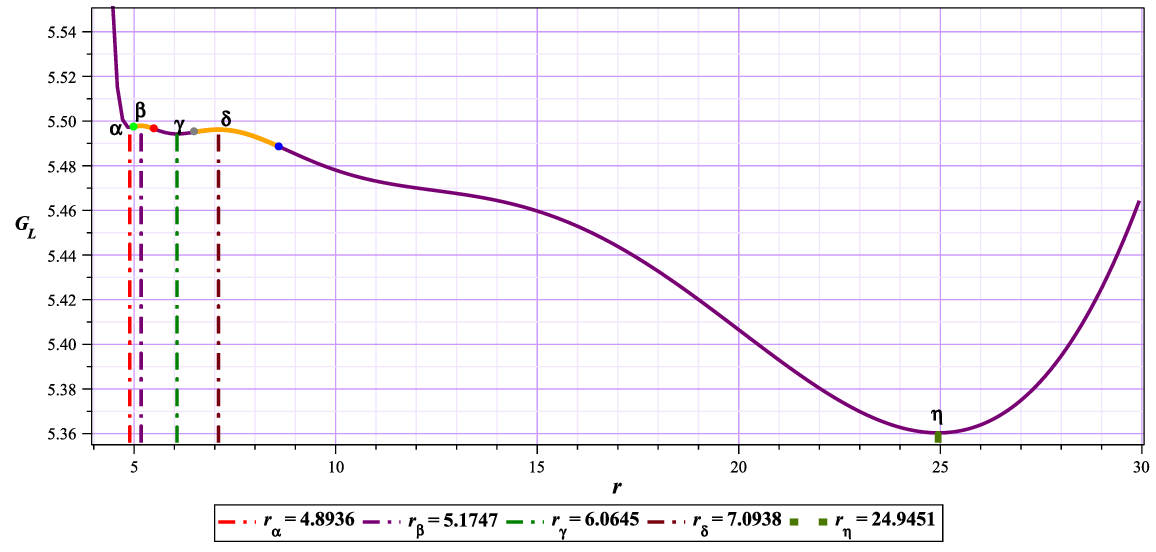}
 \label{10b}}
 \subfigure[]{
 \includegraphics[height=6.5cm,width=14.5cm]{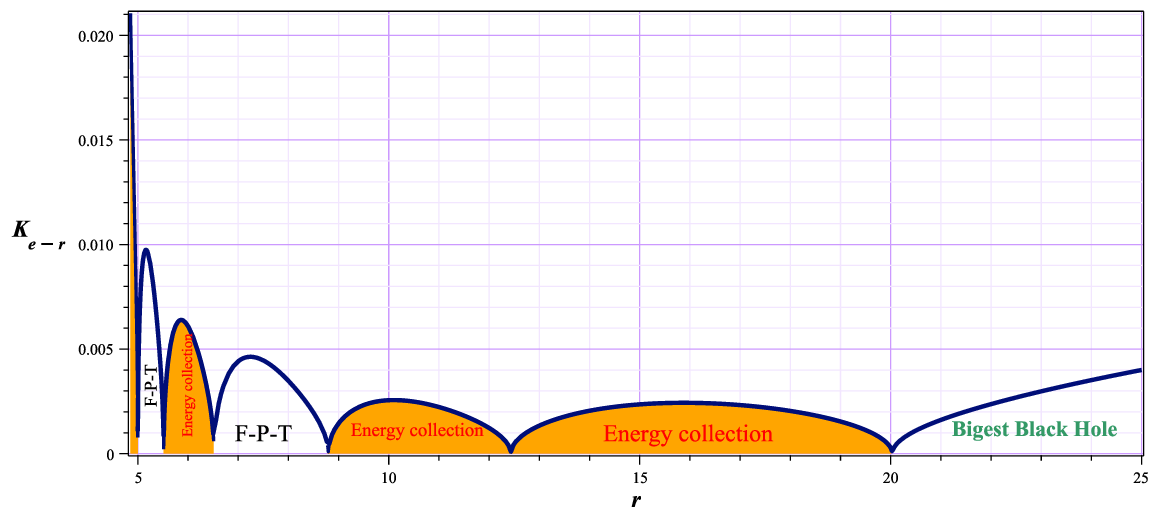}
 \label{10c}}
   \caption{\small{10a: The T diagram,\hspace{0.1cm}10b: The graph of $G_{L}$ against r for the free chosen  parameters and the FPT points (Since the green to red and gray to blue phase transitions were shown in Figures 4b and 6b, the values of them are not shown here.), \hspace{0.1cm}10c: Kramers escape rate during $\alpha\rightsquigarrow\eta$ transition with respect to $r$.}}
 \label{m10}
\end{center}
\end{figure}
In this scenario, as illustrated in Fig. \ref{m10}, the radial evolution of the model at a constant temperature exhibits only two phase transitions. Practically, the process in this case moves from the smallest  black hole to the largest stable black hole, while the intermediate black holes are gradually eliminated from larger to smaller. Although, the transition rate retains its step-like behavior (Fig. \ref{10c}), the absence of extrema in the intermediate semi-circles ensures that these regions, as previously discussed, serve as energy accumulation intervals for the eventual transfer to the largest final stable black holes.

\section{Conclusion}
In recent decades, the expanding application of classical concepts in black hole physics, combined with the fundamentally distinct structural nature of black holes, has led to continuous advancements in predicting and innovating potential behaviors and configurations of these fascinating entities.
Among various approaches, thermodynamics—due to its minimal physical and relativistic constraints—has emerged as one of the most promising frameworks for such investigations. The adoption of integrated and modern methodologies, such as topological thermodynamics or the kinetic study of black hole phase behavior via the Fokker-Planck formulation, has provided novel perspectives beyond traditional methods.\\
In this study, we aimed to analyze the kinetics and dynamic behavior of first-order phase transitions in black holes with multiple phase transitions using a particle-based approach and the Kramers escape rate.\\ Our findings indicate that the escape rate curve exhibits remarkable alignment with both the initiation and completion points of phase transitions, even in multiphase structures—consistent with prior studies \cite{55,59,60}. Based on prior investigations and the results of this study, we argue that escape rate could serve as a novel method for identifying first-order phase transitions. Specifically, if the free energy landscape within the studied domain exhibits extrema, the escape rate will accurately trace a rising-falling semicircle pattern corresponding to the transition. Conversely, if no extrema are present, the initial or final open curve along with the rising-falling semicircles  merely represent energy accumulation without an actual phase transition.\\ 
The next point we examined was that, at a specific temperature such as T = 0.0062621 which the local minima could simultaneously exist in spacetime. Since classical transitions inherently favor pathways toward lower energy states, we explored the possibility of a direct transition from the "smallest" black hole to a "large" black hole without necessarily traversing through the "small" black hole. If such a transition were feasible, we further sought to quantify its likelihood. \\
Secondly, there is usually a distance between the location of the local minimum and the start of the phase transition. Is it possible for an escape from the location of the first local minimum to the location of the second local minimum to occur directly?\\
Our analysis revealed that, based on the energy structure, such direct transitions are not possible. Instead, the model's energy configuration strictly enforces a stepwise process, maintaining sequential order without allowing for abrupt jumps or omissions. Furthermore, regarding the second question, we found that the region between the position of the local minimum and the phase transition onset serves as an energy accumulation zone, where the model retains its capacity for radial expansion without initiating a phase transition.\\
Another noteworthy observation is that in single-phase transition models, an inverse escape rate typically appears toward the end of the transition, signifying a system's tendency to finalize the process. However, in our scenario, given the sequential nature of the transitions, no inverse escapes region forms within intermediate phases.\\
Ultimately, all these findings collectively demonstrate that the Kramers particle-based approach can serve as a relatively precise tool for investigating first-order phase transitions in black holes.




\end{document}